\documentclass[a4paper,11pt]{article}
\pdfoutput=1 

\usepackage[dvipsnames]{xcolor}
\usepackage{graphicx,color}
\usepackage{jcappub}
\usepackage{enumitem}
\usepackage{braket}
\usepackage{slashed}
\usepackage{cancel}
\usepackage[utf8]{inputenc}
\usepackage{hyperref}
\usepackage[most]{tcolorbox}
\usepackage{float}
\usepackage{caption}
\usepackage{makecell}
\usepackage{epsfig}
\usepackage{multirow}
\usepackage{amsmath,amssymb}
\usepackage{tikz}
\usetikzlibrary{shapes.geometric, arrows}
\usepackage{subcaption}

\newcommand{\tti}[1]{\text{\tiny{#1}}}

\newcommand{\bv}{\Bar{v}}
\newcommand{\lr}[1]{\left(#1\right)}

\author{Filippo Revello$^{a,b}$ and Gonzalo Villa$^{c}$}

\affiliation[a]{Institute for Theoretical Physics,
Utrecht University, Princetonplein 5, 3584 CC Utrecht, The Netherlands}
\affiliation[b]{Instituut voor Theoretische Fysica, K.U. Leuven,
Celestijnenlaan 200D, B-3001 Leuven, Belgium}
\affiliation[c]{DAMTP, Centre for Mathematical Sciences, University of Cambridge, Wilberforce Road, Cambridge, CB3 0WA, UK}

\emailAdd{filippo.revello@kuleuven.be}
\emailAdd{gv297@cam.ac.uk}

\abstract{
Cosmic (super)strings offer promising ways to test ideas about the early universe and physics at high energies. While in field theory constructions their tension is usually assumed to be constant (or at most slowly-varying), this is often not the case in the context of String Theory. Indeed, the tensions of both fundamental and field theory strings within a string compactification depend on the expectation values of the moduli, which in turn can vary with time. We discuss how the evolution of a cosmic string network changes with a time-dependent tension, both for long-strings and closed loops, by providing an appropriate generalisation of the Velocity One Scale (VOS) model and its implications. The resulting phenomenology is very rich, exhibiting novel features such as growing loops, percolation and a radiation-like behaviour of the long string network. We conclude with a few remarks on the impact for gravitational wave emission.}

\usepackage{dsfont}
\title{Cosmic (super)strings with a time-varying tension}

\begin{document}

\maketitle

\section{Introduction}

Cosmic strings~\cite{Kibble:1976sj,Kibble:1984hp,Hindmarsh:1994re,Vilenkin:2000jqa,Copeland:2009ga} are among the most promising signatures of high-energy physics which could still be present in the late universe. In field theory, these strings are solitonic remnants of a broken global or gauge symmetry, with a tension $\mu$ given by the energy scale of symmetry breaking. In String Theory, they arise as fundamental objects, and their tension is related to the new fundamental scale of the theory. While initially considered as promising candidates for the seed of structure formation, the dimensionless combination $G\mu$ (where $G$ is Newton's constant) is now too severely constrained by Cosmic Microwave Background (CMB)~\cite{Charnock:2016nzm} ($G \mu \lesssim 10^{-7}$) and Pulsar Timing Arrays  ~\cite{EPTA:2023xxk,NANOGrav:2020bcs,NANOGrav:2023gor,NANOGrav:2023hvm} ($G \mu \lesssim 10^{-10}$) measurements. However, it is important to emphasize that the study of cosmic strings remains a very active topic because of their potential to seed dark matter as the QCD axion (see~\cite{Gorghetto:2018myk,Gorghetto:2020qws} and references therein), gravitational radiation (see the recent reviews~\cite{Caprini:2018mtu,Auclair:2019wcv,Sousa:2024ytl}) and other phenomenological features for which we refer to~\cite{Vilenkin:2000jqa}.

Regarding gravitational waves (GWs), it is particularly interesting that the strings are known to enter a \textit{scaling regime}\footnote{Strictly speaking, a departure from the scaling regime occurs when the equation of state of the background changes~\cite{Sousa:2013aaa,Gouttenoire:2019kij}, but we will restrict ourselves to periods with constant equation of state. In addition, we will consider scaling loosely and not take into account possible logarithmic violations due to the IR divergence of the tension of global $U(1)$ strings which is a source of disagreement in the literature, see~\cite{Correia:2024cpk,Saikawa:2024bta}. We thank Amelia Drew for discussions on this point.}, for which a network of $\mathcal{O}(1)-\mathcal{O}(10)$ super-Hubble cosmic strings contribute a constant fraction to the energy density of the background.
The scaling regime is achieved due to a balance of the energy density that enters the horizon as the Universe expands and the energy loss of the strings into small loops, which then decay predominantly\footnote{How much energy is deposited in massless radiation is another source of disagreement in the literature, see e.g.~\cite{Hindmarsh:2021mnl,Baeza-Ballesteros:2023say,Baeza-Ballesteros:2024otj}. Our main conclusions regarding scaling should not be altered by the decay channel of the loops, but the discussion on loop dynamics of section~\ref{ssc:loops} might have to be reconsidered in the future, and so would the PTA constraints (see~\cite{Kume:2024adn} for discussions on this point and~\cite{Hindmarsh:2022awe} for opportunities for multi-messenger opportunities).} to axions or gravitational waves.
Our interest in this article is the latter case, which is expected for local strings and superstrings, further described below.
Due to the scaling properties of the string configuration, the GWs sourced at every e-fold contribute a meaningful fraction to the amplitude of the GW background today, with a spectral index that depends on the equation of state of the Universe at the time the GWs are sourced~\cite{Cui:2017ufi,Cui:2018rwi} (the emission time can be inferred from the frequency of the GWs today). The GW spectrum from a network of cosmic strings therefore provides a map of the cosmological history since their formation: a golden opportunity to test the cosmology of the very early Universe.

On the other hand, one of the most generic consequences of String Theory, existence of fundamental strings aside, is the fact that there exist no free parameters, with all couplings and dimensionful quantities arising from expectation values of scalar fields - moduli - and the fundamental string scale $\sqrt{\alpha'}$. It follows that the properties of cosmic strings, whether field theoretic or fundamental, generically depend on the moduli in a stringy setting.
These scalar fields have flat potentials at tree level and their vacuum expectation values (VEVs) parametrise degeneracies in the space of solutions to the tree level equations describing the low-energy vacua of string theory.
Higher order corrections generically lift these degeneracies and result in a potential and a mass scale for the moduli, typically smaller than the string scale (see~\cite{McAllister:2023vgy} for a recent review of the problem of moduli stabilization).
It is therefore sensible to consider scenarios where the moduli are evolving on cosmological timescales, and this will invariably lead to variation of fundamental constants in the early universe, including the tension of cosmic string networks.
The GW spectrum arising from these strings will therefore be modified, providing a potential testing ground for cosmologies featuring time-varying moduli in the early Universe (see~\cite{Cicoli:2023opf} for a review). In order to make concrete statements about the GW spectrum from these strings, it is important to argue that a scaling regime is reached also when the tension varies.

In this paper we explore the phenomenology of cosmic strings with varying tension, which has already received significant attention in~\cite{Yamaguchi:2005gp,Ichikawa:2006rw,Cheng:2008ma,Wang:2012naa,Martins:2018dqg,Emond:2021vts,Conlon:2024uob}. Focusing on power-law cosmologies (with scale factor $a\sim t^{2/n}$) and a polynomially decreasing tension $\mu \sim t^{-m}$,\footnote{The $t$ denotes coordinate time in an FLRW cosmology.} we provide a comprehensive study of both the long string network and the evolution of loops, correcting some earlier claims in the literature on the former. In particular, we incorporate a time-dependent tension $\mu(t)$ in the VOS model~\cite{Martins:1996jp,Martins:2000cs}, and show how new attractor solutions can in principle arise alongside the usual scaling regime. However, only the scaling regime can be reached for realistic values of the parameters and/or the initial conditions. To further verify our conclusions, we analytically test some assumptions on the small-scale structure of the strings by solving the equations of motion for a circular loop and for the perturbations of an infinite, static string. We plan on exploiting these results to compute the GW spectrum arising from a network of cosmic superstrings with varying tension in future work~\cite{us:2024?}.

The paper is structured as follows. In Section~\ref{sc:csst}, we will briefly illustrate cosmic strings from the perspective of String Theory.
We will provide simple examples of how the tension of the strings generically depends on the moduli of the compactification, both for superstrings and ordinary, field theory strings.
In Section~\ref{sc:ct} we provide a brief summary of the dynamics of Nambu-Goto strings with constant tension, reviewing the Velocity One Scale model (VOS)~\cite{Martins:1996jp,Martins:2000cs} and the dynamics of averaged loops. 
Section~\ref{sc:tdip} contains the main results of the paper.
After writing down the microscopic equations of motion with varying tension~\cite{Emond:2021vts,Conlon:2024uob}, we provide an approximate solution for the dynamics of the circular loop, extending on \cite{Conlon:2024uob}. We appropriately generalize the VOS model and use a dynamical system analysis to argue that scaling attractor points generically exist, provided there is a condition satisfied for the so-called momentum parameter.
We find suggestive evidence that this condition is satisfied by exploring the dynamics of small perturbations to the straight, static string.
Lastly, we solve the averaged loop equations and comment on the stability of the network under percolation~\cite{Conlon:2024uob}.
Section~\ref{sec:outlook} contains our main conclusions and we provide additional details on the physics of dimensional reduction (key to understand moduli dependence in the examples we provide) in Appendix~\ref{sec:dim-red}, the calculation of the solution for the circular loop in Appendix~\ref{sec:sol-loop} and further details of the dynamical system analysis in Appendices~\ref{app:ds} and~\ref{app:k(v)}.

\section{Cosmic strings in String Theory}\label{sc:csst}

In this section we will discuss properties of cosmic strings in String Theory, distinguishing between fundamental and field theory strings. A central point (applying to both cases) is that, in a stringy context, the tension of cosmic (super)strings generically acquires a dependence in the VEVs of moduli, and can thus be dynamical. Let us first illustrate this in the case of superstrings.
In a Type II compactification without warping, the fundamental string tension can be expressed as\footnote{Although the fundamental string tension is fixed in the string frame, we will now work in the Einstein frame where the 4d (reduced) Planck mass is a fixed quantity, and the string tension can depend on expectation values of the moduli.}
\begin{equation}\label{eq:tension}
   \mu \equiv m_s^2 = \frac{g_s^{1/2} M_P^2}{4 \pi \mathcal{V}},
\end{equation}
where $g_s$ is the string coupling, and $\mathcal{V}$ is the (Einstein-frame) compactification volume in string units of $\left( 2 \pi \sqrt{\alpha'}\right)^6$. One of the main points of this work is that in String Theory all couplings are \emph{dynamical}, as they depend on the expectation values of scalar fields (the moduli), which can themselves vary with time. In the above example, the latter can be identified with the volume modulus and the dilaton, which sets the string coupling, but more exotic possibilities can also occur. This motivates us to study the dynamics of cosmic strings with a time-dependent tension, whose rich phenomenology will be analysed in the following chapters. However, we also emphasize how the results that we obtain in the rest of the paper may be used to study field theory strings with a time dependent tension, which can equally arise within string theory compactifications. In the following, we discuss both cases in detail, as well as providing some explicit examples.
This section is only meant as a general motivation towards the rest of the analysis, and the details are not important to follow the rest of the paper. The uninterested reader may skip directly to section \ref{sc:ct}.

\subsection{Cosmic superstrings}

The study of cosmic superstrings has a rich and varied history, beginning with \cite{Witten:1985fp,Polchinski:1988cn,Copeland:2003bj,Damour:2004kw,Jackson:2004zg,Copeland:2009ga}. At the microscopic level, they can be realised as either fundamental (F-)strings, or solitonic $D1$-branes (D-strings), which are dual to F-strings under S-duality.\footnote{The combination of $p$ F- and $q$ D-strings can also form stable bound states known as $(p,q)$-strings, see \cite{Witten:1985fp,Polchinski:1988cn,Copeland:2003bj,Damour:2004kw,Jackson:2004zg,Copeland:2009ga}.}
Both types can be easily produced by one of the most popular inflationary mechanisms in String Theory, $D3-\overline{D3}$ brane inflation\footnote{One can also obtain cosmic strings in 4 dimensions from a $p$-brane wrapped around a $p-1$ compact cycle. While such defects can be produced in generic brane-antibrane inflation, it is not the case for the most developed scenario of  $D3-\overline{D3}$ annihilation.}  \cite{Dvali:1998pa,Burgess:2001fx,Jones:2002cv,Sarangi:2002yt,Jones:2003da,Pogosian:2003mz,Kachru:2003sx,Burgess:2022nbx,Cicoli:2024bwq}, or more generally when the energy density of the Universe becomes order one in string units \cite{Aharonov:1987ah,Polchinski:2004ia}~\footnote{Whether the strings percolate and form a network of long, cosmic superstrings or reach thermal equilibrium instead depends on the details of the system, as explained in~\cite{Abel:1999rq,Frey:2023khe}.}. In this section, we will briefly summarize the main properties of F- and D-strings (tension, re-connection probability, etc) relevant to their cosmological evolution.

To begin with, a crucial requirement for super-strings to be cosmologically viable is a tension (string scale) that lies many orders of magnitude below the Planck scale. Indeed, even the most conservative bounds arising from CMB observations give rise to the hard limit $G \mu < 10^{-7}$ \cite{Charnock:2016nzm}, while Pulsar Timing Array (PTA) data (for example NANOGrav \cite{NANOGrav:2020bcs,NANOGrav:2023gor,NANOGrav:2023hvm}) points to even lower values of $G \mu \lesssim 10^{-11}-10^{-12}$ \cite{Blanco-Pillado:2021ygr,Ellis:2023tsl}. In heterotic string compactifications, for example, the string scale is generically of order $G \mu \sim \frac{\alpha_{\rm{GUT}}}{16 \pi} \sim 10^{-3}$, and cosmic strings are immediately ruled out \cite{Witten:1985fp}.\footnote{Moreover, various instabilities are generically present, so that a stable network cannot be formed.} To lower the string tension down to acceptable values, two main classes of mechanisms have been proposed, mainly in the context of type IIB compactifications: large extra dimensions and warping (or a combination thereof). In the following, we will review some their key properties, as well as those of the resulting cosmic superstrings.

\subsubsection{Large extra dimensions}
The first possibility is to consider a scenario where the extra dimensions are large in string units, and the string scale is diluted by the volume of the compactification manifold. Scenarios with large extra dimensions first appeared as a phenomenological solution to the hierarchy problem \cite{Arkani-Hamed:1998jmv}, and there is a natural string theory incarnation in terms of the Large Volume Scenario (LVS) \cite{Balasubramanian:2005zx,Conlon:2005ki}.
The LVS model is a scenario where all moduli can be stabilised at exponentially large values of the volume $\mathcal{V}$, allowing a very natural realisation of hierarchies between different scales. As an example, one can easily see from \eqref{eq:tension} that exponentially large volumes can introduce a significant separation between the 4d Planck scale and the string scale, allowing one to bypass the bounds discussed in the previous section on the tension of cosmic strings. Realistic values of the stabilised volume in LVS are of order $ \mathcal{V} \sim 10^6 -  10^{13}$ (and $g_s \sim 0.1)$, corresponding (from \eqref{eq:tension}) to a string tension in the range $G \mu \sim 10^{-9}-10^{-16} $, which can easily evade observational bounds. 

However, as emphasized in \cite{Conlon:2022pnx}, this need not be the case during the entire history of the universe, as such constraints only apply from the CMB emission time onwards. It is indeed possible to have a string scale that is marginally lower then the 4d Planck scale in the very early universe (say immediately after inflation), which is subsequently lowered by the rolling of the volume modulus along its potential.\footnote{In principle, the dilaton could also be rolling along its potential and give rise to a decreasing tension. However, the most commonly studied scenarios of moduli stabilization (such as LVS and KKLT) are based on the GKP construction \cite{Giddings:2001yu}, where the axio-dilaton is stabilized at much higher scales together with the complex structure sector.} Given the presumed hierarchy between the electroweak and inflationary scales, $\Lambda_{{\rm EW}} \ll \Lambda_{{\rm inf}}$, this is not only possible but actually a very desirable feature of a realistic model. In the context of the LVS scenario such a cosmology has been studied in detail \cite{Conlon:2022pnx,Apers:2022cyl,Revello:2023hro,Apers:2024ffe,Conlon:2024hgw,Conlon:2024ene,Apers:2024dtn}, consisting in a succession of different kination and tracker epochs mediated by the volume (see also the earlier work \cite{Conlon:2008cj}). 

The presence of large extra dimensions also reduces the probability\footnote{Low intercommutation probability is typically assumed to be a property of superstrings. However, it has been argued that flux tubes of a non-abelian dark sector could have similar properties~\cite{Yamada:2022aax,Yamada:2022imq}.} for strings to scatter off each other, as they can now miss each other in the extra dimensions. If the strings were truly able to propagate within the extra-dimensional bulk, this would lower the reconnection probability by a factor of the volume, and quickly reach very small values. However, the position of the strings along the extra dimensions is a worldsheet scalar and corrections to e.g. the flat space metric will stabilize it.
Using this logic, it has been argued~\cite{Jackson:2004zg} that warping effects (generically arising due to the presence of sources of energy along the extra dimensions) localize the strings in a small extra-dimensional region, yielding a reconnection probability of
\begin{equation}
   P \simeq 0.5 g_s^2 \left[\frac{6 \pi}{\log \left( A \mathcal{V}\right)} \right]^3.
\end{equation}
with $A$ a model-dependent, order one number (at weak warping). This depends very weakly in the volume.
It is worth noting that brane-antibrane inflation, which naturally produces cosmic superstrings at its endpoint, has recently been reasessed~\cite{Burgess:2022nbx,Cicoli:2024bwq} and a potential with a hierarchical separation between the inflationary and late-time scales has been obtained.
The volume modulus would evolve from its value during inflation to its exponentially large value at the true, late-time minimum, providing a concrete example of the ideas motivating this article.

\subsubsection{Warped throat}

The second possibility is to localise the strings at the bottom of warped throat geometry attached to a ``bulk" Calabi-Yau geometry, so that the original string tension is redshifted by an amount which can be tuned exponentially, depending on the geometric parameters. In particular, the latter can be described by a metric of the form~\cite{Dasgupta:1999ss,Giddings:2001yu}
\begin{equation}\label{eq:amb-metric}
    d s^2=e^{2 A(y)} g_{\mu \nu} d x^{\mu} d x^{\nu} +e^{-2 A(y)} g_{m n} d y^n d y^m,
\end{equation}
where $A(y)$ is the warping factor. The tension for a string localised at $y=y_0$ can be written as
\begin{equation}
   \mu= m_s^2 = e^{4A(y_0)}\frac{g_s^{1/2}M_P^2}{4 \pi \mathcal{V}},
\end{equation}
where the volume is not assumed to be exponentially large (although one must still work in the large volume limit to apply the supergravity approximation). The interest here lies in the fact that brane-antibrane inflation relies on a deep throat to obtain slow-roll (see~\cite{Baumann:2014nda} for extensive discussion on the topic, and~\cite{Burgess:2022nbx,Cicoli:2024bwq} for recent work).
This is also a crucial ingredient in the one of the most well known (and debated) constructions of de Sitter vacua in String Theory, known as KKLT \cite{Kachru:2003aw}. In particular, a string theoretic embedding of brane inflation within string compactifications in the KKLT framework was achieved in \cite{Kachru:2003sx}.
The vanilla KKLT does not easily allow for modifications to the standard cosmological history and a time-dependent string tension through a rolling volume. This is simply because there is not enough space for the volume to roll before it hits the minimum of its potential. Although we will not pursue this further, let us however notice that there is in principle a different possibility for a time-dependent tension in this case. There is a new modulus, the conifold, which determines the magnitude of the warping factor.\footnote{Post-inflationary dynamics of this modulus leading to a Hagedorn phase were considered in~\cite{Frey:2005jk}.}

\subsection{Field theory strings in String Theory}

Let us now illustrate how, within the framework of String Theory, the tension of ordinary field theory strings can also depend on time. Field theory cosmic strings arise as solitonic solutions to the equations of motions in a variety of QFTs, and here we will present a few explicit realisations. 

\subsubsection{Local $U(1)$ strings}
Let us consider the simple example of a $U(1)$ gauge theory which is spontaneously broken.
The tension is given by the VEV of the Higgs field that breaks the symmetry in the vacuum.
This VEV, however, is an energy scale which in string theory must either be the fundamental string scale or arise \textit{dynamically}.
To see this, focusing in IIB compactifications (where gauge fields arise as open strings stretching between branes - see~\cite{Ibanez:2012zz,Marchesano:2022qbx,Marchesano:2024gul} for reviews), notice that the relevant part of the action for the worldvolume theory of a Dp-brane has the schematic form
\begin{equation}
    S_{\text{DBI}}=T_p\int{d^{p+1}x \, \sqrt{h} \lr{-\frac{1}{4}F_{AB}F^{AB}- D_\mu \phi D^{\mu} \phi^*-V(\phi, \phi^*)}} \, .
\end{equation}
Here $T_p \sim \alpha'^{-(p+1)/2}$ and $h_{AB}$ is the pullback of the ambient metric~\eqref{eq:amb-metric} on the worldvolume of the brane.
To illustrate the point, let us consider a toy model D3-brane with a worldvolume abelian-Higgs model:
\begin{equation}
    V(\phi,\phi^*)=\frac{\lambda}{4}
    \lr{|\phi|^2-v^2}^2\, ,
\end{equation}
It is well known that this theory admits cosmic string solutions arising from the breaking of the U(1) gauge symmetry, and we intend to argue that the dimensionless tension $G\mu$ of these strings is moduli-dependent even if the parameter $v$ was not (noting that in a concrete construction it would be).
The reason is that gravitational processes generically inherit moduli dependence from, at least, the compactification volume.
We provide a short summary of the physics of dimensional reduction in Appendix~\ref{sec:dim-red}, and simply note here that in Einstein frame the tension of a D3-brane in terms of the Einstein-frame volume modulus $\mathcal{V}$ reads:
\begin{equation}
    T_3=\frac{M_p^4}{8\pi \mathcal{V}^2}\, .
\end{equation}
Thus, upon canonical normalization:
\begin{equation}
    S_{cn}=\int{d^3 x \sqrt{h}\lr{-\frac{1}{4}F_{\mu \nu}F^{\mu \nu}-D_\mu \Tilde{\phi} D^\mu \Tilde{\phi}^*-\frac{\Tilde{\lambda}}{4}\lr{ \Tilde{\phi}^2-\Tilde{v}^2}^2}}\, .
\end{equation}
where we have redefined the parameters to
\begin{equation}
\Tilde{\phi}=\phi \sqrt{T_3}\, \qquad \Tilde{\lambda}=\lambda /T_3\, , \qquad \Tilde{v}=v\, \sqrt{T_3}\, ,
\end{equation}
(the gauge field also needs to be redefined but this is not important for the present discussion). The effect of this field redefinition is to introduce a moduli dependence on the scalar field vev, which is further inherited by the string tension. In particular, the latter is given by
\begin{equation}
    \mu=2\pi \Tilde{v}^2=\frac{(v\,\cdot M_p)^2}{4\mathcal{V}^2}M_p^2\, .
\end{equation}
We thus conclude that the tension of the string (in Planck units) generically features a dependence on the volume modulus of the compactification, inherited from the dependence of the Planck scale in this modulus.
In realistic scenarios, this scalar must acquire a mass and develop a constant vacuum expectation value, at least, before Big Bang Nucleosynthesis, since time-variations of the Planck scale from this time are very constrained (see the discussion in~\cite{Cicoli:2023opf}).
However, in concrete constructions these moduli are generically very light with respect to the string scale and so it is plausible that they might have been cosmologically active.

An additional example of a scalar which could have been cosmologically active at late times whilst changing the tension of a network of cosmic strings and being experimentally viable is the volume of a 4-cycle that makes a small contribution to the overall volume of the compactification manifold (and so to the Planck scale). Let us assume that a hidden D7-brane wraps this cycle and features an abelian-Higgs model. Then, on top of the above, four-dimensional physicists observe an additional dependence in the volume $\tau$ of the 4-cycle that is wrapped. Indeed, at energies much lower than $(2\pi \sqrt{\alpha'})^{-1}\tau^{-1/4}$, as described in Appendix~\ref{sec:dim-red}, the effective description of the physics arises from the dimensional reduction of the worldvolume action, which for the present discussion yields an effective brane tension $T_7 \tau (2\pi \sqrt{\alpha'})^4$.
The discussion follows as above changing $T_3 \to T_7 \tau (2\pi \sqrt{\alpha'})^4$, and so an additional dependence in this modulus appears.

\subsubsection{Axion strings}

A slightly different class of examples is provided by axion strings, which can arise in the presence of compact scalars. Such defects are characterised by the fact that as one completes a full circle around the string core in spacetime, the field winds an integer number of times around its compact domain. Conventionally, axion field ranges are measured in terms of the axion decay constant $f_a$, which determines the periodicity of the axion field as $a \rightarrow a + 2 \pi f_a$. The calculation of the axion string tension contains two contributions. The first is a model-independent, IR-divergent term which arises from the energy density of the field configuration outside the string core. It can be estimated as
\begin{equation}
    T_{IR} \simeq \pi f_a^2 \log \left( \frac{r_{\text{IR}}}{r_{\text{core}}}\right),
\end{equation}
where $r_{\text{core}}$ is the core radius (of order $m_s^{-1}$ in string theory examples) and $r_{\text{IR}}$ an IR cutoff, usually taken to be the distance to the next string in the network. The second contribution comes from the string core, which is a singular object from the EFT perspective, and carries information about the UV completion (see \cite{Lanza:2020qmt,Lanza:2021udy}). In the case of string theory axions, the core singularity is resolved in terms of fundamental strings or branes wrapping an internal cycle of the compactification manifold. Therefore, the tension will be equal to the product of a $p+1-$brane tension and the volume of the $p-$cycle it is wrapping in the extra dimensions:
\begin{equation}\label{eq:axion-tension}
    T_{UV} = T_{p+1} \,  {\rm{Vol}\Sigma_p} =  \frac{ g_s^{\frac{p-3}{4}}}{2^{\frac{p+1}{2}} (2 \pi)^{\frac{p-1}{2}}} \frac{M_P^{p+1}}{\mathcal{V}^{\frac{p+1}{2}}} {\rm{Vol}\Sigma_p}
\end{equation}
As in the case of $U(1)$ strings, this depends on the string coupling and overall volume, as well as the moduli controlling the size of $\Sigma_p$.

Alternatively, one can also see this from a lower-dimensional perspective, where the tension is controlled by the decay constant $f_a$. In the context of 4d, $\mathcal{N}=1$ supergravity, axions are paired to a geometric modulus within a chiral multiplet
\begin{equation}
    T= s+ i a.
\end{equation}
In string theory compactifications, axions arise from the dimensional reduction of higher forms, and inherit a discrete shift symmetry from the higher dimensional gauge symmetry \cite{Witten:1984dg,Svrcek:2006yi}. As an example, one may consider $C_4$ axions \cite{Conlon:2006tq,Cicoli:2012sz} in type IIB flux compactifications, obtained by integrating the $C_4$ form on 4-cycles $\Sigma_i$ of the compactification manifold:
\begin{equation}
c_i=\frac{1}{(2 \pi)^2 \alpha'^2} \int_{\Sigma_i} C_4.
\end{equation}
They are paired to $4-$cycle (Einstein frame) volumes as $T_i = \tau_i +i c_i$, where
\begin{equation}
    \tau_i=\int_{\Sigma_i} e^{-\phi} \, \rm{d} Vol.
\end{equation}
Their K\"ahler potential is of the schematic form\footnote{The overall K\"ahler potential for K\"ahler moduli is $K = - 3 \log \mathcal{V}$, with an overall volume $\mathcal{V}$ that can be written in terms of 4-cycle volumes. Here we are keeping the other moduli fixed and neglecting mixing.}
\begin{equation}
    K = - k \log \left( T+\bar{T} \right) - \log (S+\bar{S})
\end{equation}
where $S= C_0+i e^{-\phi}$ is the axio-dilaton, and $k$ a constant. Similar expressions also hold for other even axions, or odd form axions in type IIA compactifications. The kinetic term then reads
\begin{equation}
    \mathcal{L} \supset \frac{k^2}{4}\frac{\partial_{\mu}\tau \partial^{\mu}\tau}{\tau^2}+\frac{k^2}{4}\frac{\partial_{\mu}a \partial^{\mu}a}{\tau^2},
\end{equation}
from which one can read off an axion decay constant that depends on the vacuum expectation value of the (canonically normalised) K\"ahler modulus $\varphi$ as
\begin{equation}
    f_a = M_P \,  e^{-\frac{\sqrt{2}}{k} \langle \varphi \rangle}, \quad \quad \text{where} \quad  \varphi = \frac{k}{\sqrt{2}} \log \tau.
\end{equation}
It follows that both contributions to the axion string tension, $T_{IR}$ and $T_{UV}$, can vary with the moduli. Indeed, it was recently argued in \cite{Reece:2024wrn} (See also \cite{Benabou:2023npn}) that the core tension of a fundamental axion string can be estimated from the decay constant as\footnote{In any case, the tension must satisfy the bound $T_{UV} \leq 2 \pi M_p f_a$ to be consistent with the axion Weak Gravity Conjecture (aWGC) \cite{Arkani-Hamed:2006emk}.}
\begin{equation}
    T_{UV} = 2 \pi S_{\rm{inst}} f_a^2
\end{equation}
in a variety of examples (reproducing e.g. the scalings in \eqref{eq:axion-tension}), where $S_{\rm{inst}}$ is the action of the instanton generating the axion potential. We thus conclude that both superstrings as well as field theory (global and local) strings inherit dependence in moduli in a stringy setting, which potentially translates to (space)time dependence if the latter have non-trivial profiles.

\section{String networks with constant tension}\label{sc:ct}

In this section, we introduce the scaling regime and outline the effects of a reduced inter-commutation probability in the energy density of a scaling cosmic string network with constant tension.
This section is mostly review material but will be useful later in section~\ref{sc:tdip} where we will generalize the scenario.
It is well known that, modulo the subtleties discussed in the introduction, cosmic string networks reach a configuration dubbed as the scaling regime \cite{Kibble:1984hp,Bennett:1985qt,Austin:1993rg} (see also \cite{Vilenkin:2000jqa} for a pedagogical discussion): a self-similar configuration in which the energy density $\rho$ of the system remains a fixed fraction of the energy density of the background (which we implicitly assume is much larger than that of the strings, thus leading the expansion of the Universe). 

\subsection{Equations of motion}\label{ssc:EOMtc}

 We begin by recalling a few basic facts on the dynamics of strings in an expanding spacetime. This will be useful for the generalisation to the case of a time dependent tension in Section \ref{sc:tdip}, and to provide some definitions. In our notation, the string worldsheet is parametrised by the variables $\zeta = (\zeta^0,\zeta^1)$, while the spacetime coordinates denoted by $X^{\rho}(\zeta)$. The worldsheet metric is given by
\begin{equation}
\gamma_{a b}= \frac{\partial X^{\mu}(\zeta)}{\partial \zeta^a} \frac{\partial X^{\nu}(\zeta)}{\partial \zeta^b} g_{\mu \nu}, \quad \quad \text{where} \quad \quad g_{\mu \nu} = a^2(\tau) \eta_{\mu \nu} 
\end{equation}
is the form of the target space metric for an FRW cosmology (switching to conformal time $\tau$). The effective dynamics of such strings are described by the usual Nambu-Goto action, of the form
\begin{equation}\label{eq:NGc}
    S=- \mu \int d^2 \zeta\,   \sqrt{-\gamma}.
\end{equation}
Performing a variation with respect to the action, one obtains the equation of motion
\begin{equation}\label{eq:eomc}
    \frac{1}{ \sqrt{-\gamma}} \partial_a \left(  \sqrt{-\gamma} \gamma^{ab} g_{\rho \sigma} X^{\sigma}_{\, ,b} \right)-\frac{1}{2} \partial_{\rho} g_{\lambda \sigma} \gamma^{a b} X^{\lambda}_{\, ,a} X^{\sigma}_{\, ,b} =0,
\end{equation}
 At this point, we are free to make a gauge choice to simplify the expressions, and choose a gauge where $\dot{X} \cdot X' =0$ (equivalently, $\gamma_{0 1}=\gamma_{10}=0$) and worldsheet time is identified with the conformal time coordinate, \emph{i.e.} $X^0(\zeta) = \tau$. In this gauge, we will be denoting spacetime coordinates as $X^{\mu}= \left(\tau,x^i\right)$, with $x^2 \equiv x^i x_i$. Upon defining the energy density parameter
\begin{equation}\label{eq:eps}
    \varepsilon \equiv \sqrt{\frac{x'^2}{1-\dot{x}^2}},
\end{equation}
the equation of motion \eqref{eq:eomc} can be recast as
\begin{equation}\label{eq:E0c}
    \dot{\varepsilon}+  2 \frac{\dot{a}}{a} \varepsilon \dot{x}^2 =0
\end{equation}
 for $\rho = 0$, and 
\begin{equation}\label{eq:Eic}
    \ddot{x}_i+ 2 \frac{\dot{a}}{a} (1-\dot{x}^2) \dot{x}_i- \frac{1}{\varepsilon} \left( \frac{x_i^{'}}{\varepsilon} \right)^{'} =0.
\end{equation}
for $\rho=i$. Notice that the dots here denote
derivatives with respect to conformal time $\tau$, not to be confused with coordinate time $t$. To go from a microscopic to a coarse-grained description of the network, a key step is to integrate over the spatial coordinate of each string and consider them as point-like objects. To this purpose, the total energy and average squared velocity of a string can be defined as
\begin{equation}
    E = \mu(\tau) a(\tau) \int \varepsilon d \zeta^1 \quad \quad \text{and} \quad \quad v^2= \frac{\int d \zeta^1 \,\varepsilon \dot{x}^2 }{\int  d \zeta^1 \, \varepsilon }.
\end{equation}
From the fact that the energy density of the strings will scale as $\rho \sim E /a(\tau)^3$, one can derive macroscopic equations as shown in the next section.

\subsection{VOS model}

The scaling behaviour mentioned in the first paragraph is well captured by the VOS model ~\cite{Martins:1996jp,Martins:2000cs}, which describes the evolution of a string network in terms of a characteristic length $L$, defined through
\begin{equation}\label{eq:network-density}
    \rho=\frac{\mu}{L^2}\, ,
\end{equation}
where $\mu$ is the string tension and $v$ the rms string velocity defined above. $L$ also admits an interpretation as the average distance between strings, and is sometimes called the correlation length. 
The two fundamental equations in the VOS model are essentially integrated versions of the microscopic equations of motion for an ideal (Nambu-Goto) string, \eqref{eq:E0c}- \eqref{eq:Eic}, supplemented with a loop-chopping parameter $\Tilde{c}$ which captures the energy loss of the network into loops.
The equations read
\begin{equation}\label{eq:VOSmu}
   \frac{{\rm d} v}{{\rm d}t} = (1-v^2) \left[\frac{k(v)}{L}-2 H v \right]
\end{equation}
and
\begin{equation}\label{eq:VOSL}
    2 \frac{{\rm d} L}{{\rm d}t}= 2 H L(1+v^2)+\tilde{c}v.
\end{equation}
In the scaling regime (which is an attractor solution of the equations above), the characteristic length $L$ evolves as
\begin{equation}\label{eq:vos1}
 L=\xi t\,     \qquad \quad \xi= \sqrt{ \frac{n}{8}\frac{k(\bar{v})(k(\bar{v})+\tilde{c})}{1-\frac{2}{n}}},
\end{equation}
where $n$ characterises the cosmological period through $\rho\sim a^{-n}$ (\textit{i.e:} $a(t) \sim t^{2/n}$) and the important quantity $c$ quantifies the efficiency of energy loss of the network (more on this later). Following~\cite{Cui:2018rwi}, the root mean squared velocity $\bv$ is given, for our purposes, by the expression 
\begin{equation}\label{eq:vos2}
    \bv=\sqrt{\frac{n}{2}\frac{k(\bv)}{k(\bv)+\tilde{c}}\lr{1-\frac{2}{n}}}\,.
\end{equation}
The function $k(v)$ has the microscopic definition
\begin{equation}
    \left\langle\left(1-\dot{\mathbf{x}}^2\right)(\dot{\mathbf{x}} \cdot \hat{\mathbf{u}})\right\rangle \equiv k v\left(1-v^2\right),
\end{equation}
where $\hat{\mathbf{u}}$ is the unit curvature vector of the string. It is known in the literature as the momentum parameter,
and encodes information about small scale structure of the string. A phenomenologically motivated expression is \cite{Martins:2000cs}
\begin{equation}\label{eq:k(v)}
    k(v)=\frac{2 \sqrt{2}}{\pi}\left(1-v^2\right)\left(1+2 \sqrt{2} v^3\right) \frac{1-8 v^6}{1+8 v^6},
\end{equation}
which reproduces the correct non-relativistic limit and vanishes for the flat spacetime value of $v^2 = \frac{1}{2}$. For an expanding spacetime, the rms string velocity will asymptote to a value just below $v^2 < \frac{1}{2}$\cite{Vilenkin:2000jqa}, so that $k(\bar{v})$ is small and positive in accordance with \eqref{eq:vos1}-\eqref{eq:vos2}.
In upcoming sections we will argue that if the tension decreases sufficiently fast with respect to the expansion of the Universe, a scaling regime is achieved with $\bar{v}^2>1/2$ and $k(\bar{v})<0$.
Before concluding the section, we note that the scaling behaviour has been also studied in the context of cosmic superstrings~\cite{Sakellariadou:2004wq,Avgoustidis:2005nv,Pourtsidou:2010gu,Avelino:2012qy,Sousa:2013aaa,Sousa:2016ggw}, where the reconnection probability $P$ is smaller than one.
The qualitative features of the discussion are not altered and a scaling regime is achieved, albeit the different quantities ($c,\xi$, etc) inherit a $P$-dependence. In the following we will simply assume the VOS model.

\subsection{Loop decays}
In order to maintain the scaling regime, the network must lose energy at the same rate as the background.
This is done by emission of loops.
These loops then decay into axions, massive matter or gravitational waves, depending on the origin of the defect.
We mentioned the discrepancies in the literature regarding these decay channels on the introduction, and here we will focus in local, Nambu-Goto strings.
Simulations \cite{Blanco-Pillado:2013qja} suggest that around $90\%$ of the energy of the network is released into small ultra-relativistic loops that are red-shifted away. The other fraction $\mathcal{F}_\alpha \simeq 0.1$ of the energy is deposited into small loops of length that is a fraction $\alpha$ of the correlation length:
\begin{equation}
    \ell(t_i)=\alpha L(t_i)= \alpha \xi t_i \, ,
\end{equation}
where in the second equality we have assumed the scaling regime of the VOS. 
Local cosmic string loops decay through gravitational radiation.\footnote{Global strings can also decay by radiating massless Nambu-Goldstone bosons, with $\frac{d E_{\tti{GW}}}{dt} \propto v^2$, where $v$ is the vev of the corresponding scalar field.}
In the frequency regime of interest, the power of emission is independent of their length, and reads
\begin{equation}\label{eq:loop-gw-emission}
    \frac{d E_{\tti{GW}}}{dt}=\Gamma G \mu ^2\, ,
\end{equation}
where $\Gamma\simeq 50$ is determined by simulations (for which we assume the results to hold for superstrings with varying string tension on the grounds of adiabaticity), and the energy of the loop is $E=\mu \ell$.
Thus, with a constant string tension, the temporal evolution of one string loop is:
\begin{equation}\label{eq:loop-ev-constant-tension}
    \ell(t)=\alpha  t_i-\Gamma G\mu (t-t_i)\, .
\end{equation}
Let us assume that the network loses energy at a rate \cite{Kibble:1984hp} 
\begin{equation}\label{eq:eloss}
    \frac{d \rho}{dt}=\tilde{c} \bv \frac{\rho}{L}\, ,
\end{equation}
where $\tilde{c}$ is known as the loop chopping efficiency. Assuming that a fraction $\mathcal{F}_\alpha$ of this energy goes into loops of length $\ell(t_i)$, one obtains the following loop production rate
\begin{equation}\label{eq:lprod}
    \mathcal{F}_\alpha\frac{d \rho}{dt}=\mu \gamma \ell \frac{dn}{dt} \to \frac{dn}{dt}=\frac{\mathcal{F}_\alpha}{\gamma}\frac{\tilde{c} \bv}{\alpha}\frac{1}{L^4}\,.
\end{equation}    
where $\gamma$ is a Lorentz factor which takes into account that $\ell$ is the proper length of the loop, while the energy is computed in the rest frame of the mother string.
The loop production rate during scaling therefore features a behaviour $dn/dt=A/t^4$ during scaling, where $A$ depends on the background~\cite{Blanco-Pillado:2017oxo,Blanco-Pillado:2013qja} and can be inferred from simulations.

\section{The effect of a time-dependent tension}\label{sc:tdip}

The results we reviewed in the previous section only apply to the case of a constant string tension. However, as we have seen in Section \ref{sc:csst}, String Theory motivates scenarios where the tension is time-dependent.
The canonical example are models with large extra dimensions, where the fundamental string tension  (neglecting warping) is related to the internal volume as
\begin{equation}
   \mu=  m_s^2 = \left(\frac{M_P}{4 \pi } \right)^2 \frac{\sqrt{g_s}}{\mathcal{V}} ,
\end{equation}
and the extra-dimensional volume $\mathcal{V}$ is measured in stringy units of $(2\pi \sqrt{\alpha'})^6$. In phenomenologically interesting scenarios, such as volume kination or tracker epochs \cite{Apers:2024ffe}, the volume can vary with time, and thus induce a time-dependent string tension.
In both cases the tension evolves as $\mu \sim t^{-m}$, with $m=1$ and $m=2/3$ for kination and the tracker respectively.
The goal of this section will be to extend the original VOS model with a time-dependent string tension, and also discuss the evolution of loops\footnote{\label{ft:eoms} An early attempt to study the VOS with varying tension~\cite{Yamaguchi:2005gp} imposed the string tension to depend on coordinate, rather than worldsheet time, in the Nambu-Goto action. This breaks worldsheet reparametrisations and leads to the wrong equations of motion (see also \cite{Emond:2021vts,Conlon:2024uob}). It would be interesting to adapt the methods of~\cite{Martins:2018dqg} to the equations presented in this article.}.
We shall see that allowing for a varying tension entails a rich phenomenology, making scaling far from obvious.
The goal of this article is to argue that a scaling solution is nevertheless achieved, and we will make some comments on the timescales for its stability to the percolation phenomenon recently discussed in~\cite{Conlon:2024uob}.
Towards the end of the section we will also extend the analysis to the population of loops, whose distribution is crucial for the computation of gravitational wave spectra.

\subsection{Equations of motion and analytical aspects}\label{ssc:tVOS}

The equations of motion obtain in Section \ref{ssc:EOMtc} will clearly need to be modified in case of a time-dependent tension.
The effective dynamics of the strings will now be described  by a modified Nambu-Goto action, of the form
\begin{equation}\label{eq:NG}
    S=- \int d^2 \zeta\,  \mu \left(X^{\nu}(\zeta)\right)\sqrt{-\gamma}.
\end{equation}
Performing a variation with respect to the action, one obtains the equation of motion
\begin{equation}\label{eq:eomvt}
    \frac{1}{\mu \left(X^{\nu}\right) \sqrt{-\gamma}} \partial_a \left( \mu \left(X^{\nu}\right) \sqrt{-\gamma} \gamma^{ab} g_{\rho \sigma} X^{\sigma}_{\, ,b} \right)-\frac{1}{2} \partial_{\rho} g_{\lambda \sigma} \gamma^{a b} X^{\lambda}_{\, ,a} X^{\sigma}_{\, ,b} -\frac{ \partial_{\rho} \mu \left(X^{\nu}\right)}{\mu \left(X^{\nu}\right)}=0,
\end{equation}
where the dependence on $X^{\nu}(\zeta)$ is left implicit. Eq. \eqref{eq:eomvt} was previously derived in \cite{Emond:2021vts,Conlon:2024uob}, and differs from the equation in \cite{Yamaguchi:2005gp} only by the last term, which arises precisely through the dependence of the tension on the embedding coordinates. From now onwards, we also assume that the tension only depends on time, that is $\mu \equiv \mu \left(X^0(\zeta)\right)$. As in the case of constant tension, we choose a gauge where $\dot{X} \cdot X' =0$ (and $\gamma_{0 1}=\gamma_{10}=0$), and identify conformal and worldsheet time, $X^0(\zeta) = \tau$. Notice, however, that this is only possible \emph{after} the equations of motion have been written down out, as doing so in the action would have lead to a wrong result for \eqref{eq:eomvt}. Defining the energy density of a string as in \eqref{eq:eps}, the equation of motion \eqref{eq:eomvt} for $\rho = 0$ obtains an additional term with respect to the constant tension case
\begin{equation}\label{eq:E0vt}
    \dot{\epsilon}+ \left(2 \frac{\dot{a}}{a}+\frac{\dot{\mu}}{\mu}\right) \varepsilon \dot{x}^2=0.
\end{equation}
On the other hand, the equations of motion for $\rho=i$ become
\begin{equation}\label{eq:Eivt}
    \ddot{x}_i+ \left(2 \frac{\dot{a}}{a}+ \frac{\dot{\mu}}{\mu} \right) (1-\dot{x}^2) \dot{x}_i- \frac{1}{\varepsilon} \left( \frac{x_i^{'}}{\varepsilon} \right)^{'} =0,
\end{equation}
where we are again denoting derivatives with respect to conformal time $\tau$ with a dot. Notice how, in both cases, the equations of motions are equivalent to those of a string with a constant tension, and a modified conformal Hubble rate (denoted as $\mathcal{H})$ \footnote{This is only true at the level of the microscopic equations of motion, as additional factors of $H$ appear after averaging.}
\begin{equation}\label{eq:H}
   2 \mathcal{H}' = 2 \mathcal{H} + \frac{\dot{\mu}}{\mu}.
\end{equation}
At the microscopic level, the effects of the expansion of the Universe and those of varying the tension play the same role: they induce a friction term of either sign.
In particular, in the Minkowskian case $\mathcal{H}=0$ with decreasing tension, the strings tend to grow to preserve energy.

The interplay between varying tension and Hubble expansion leads to a rich phenomenology.
For instance, as explained in \cite{Emond:2021vts}, for the special case $\mu \sim a(t)^{-2}$ the action becomes scale-invariant, and the equations of motion reproduce those of a constant tension string in flat space-time. Even more striking is the case when the RHS of \eqref{eq:H} is negative, since the loops grow faster than the scale factor \cite{Conlon:2024uob}; the strings behave as if they were placed in a contracting universe.
It is this case that we intend to study in this article\footnote{Note that the universe is still undergoing expansion, so the dynamics of the network will be different to that of cosmic strings in an actual contracting universe, studied in \cite{Avelino:2002hx,Avelino:2002xy}.}.

\subsubsection{The circular loop}
To be more concrete about our claims, let us find a solution for the microscopic equations of motion in an arbitrary power-law cosmology $a\sim t^{2/n}$ with $\mu \sim t^{-m}$.
Note that this includes the constant tension case $m=0$, but we are not aware of this solution having been reported elsewhere in the literature (although its qualitative features are understood).
It is well known that, when $\textbf{x}'\cdot \dot{\textbf{x}}=0$, the equations of motion in Eqs.~\eqref{eq:E0vt} and~\eqref{eq:Eivt} are not independent.
To solve the condition $\textbf{x}'\cdot \dot{\textbf{x}}=0$ we focus on the circular loop:
\begin{equation}
\textbf{x}=
R(\tau)L(\cos (\zeta^1/L),\sin (\zeta^1/L),0)\, ,
\end{equation}
where $L$ is an arbitrary dimensionful scale.
It is well known that the circular loop in Minkowski space satisfies $R(t)=\cos (\tau/L+\alpha)$, for arbitrary $\alpha$.
In Appendix~\ref{sec:sol-loop} we extend this solution for arbitrary power-law cosmologies.
The main features of the solution are as follows.
The function $R(\tau)$ satisfies
\begin{equation}\label{eq:r-sol-exp}
R(\tau)=e^{-q\int{\frac{d\tau}{\tau}\sin (f(\tau/L))^2}}\cos (f(\tau/L))\, ,
\end{equation}
where $q=(4/n-m)/(1-2/n)$ and $f(\tau/L)$ is a function which, in coordinate time, interpolates between $f(t/L)\sim a(t)^2/t^m$ in the superhorizon limit $t/L\ll 1$ and $f(t/L)\sim (t/L)^{1-m/2}$ in the subhorizon limit.
It is worth noting that $R(\tau)$ is an oscillating function and its maximum amplitude behaves as
\begin{equation}
R_{max}(\tau) \sim \lr{\frac{\tau}{\tau_0}}^{-q/2}\, ,
\end{equation}
together with a proportionality constant that depends  on the initial conditions and an oscillating function that relaxes to one.
In the case with constant tension, the amplitude radius therefore grows as $R(\tau)\sim 1/a(t)$, namely the physical radius remains constant, and as the tension varies we have $R(t) a(t) \sim t^{m/2}$, so the physical radius grows.

\subsubsection{The momentum parameter: analytical modelling}\label{sec:momentum-parameter}

The momentum parameter $k(v)$ plays a crucial role in the averaged equations of motion, and in this section we would like to gain some understanding on its behaviour from a microscopic point of view.
As will be clear in the next section, a crucial aspect of achieving a scaling regime (when $m\, n>4$) is that the momentum parameter switches sign when crossing the threshold $v^2=\frac{1}{2}$.
Let us provide some evidence in support of this claim. To do so, we study the evolution of small scale structure with varying tension.
Let us follow~\cite{Vilenkin:2000jqa} and study linear perturbations with respect to the static, straight string:
\begin{equation}
    \textbf{x}=\textbf{c} \zeta^1\, ,
\end{equation}
which can be thought of as the $L\to \infty$ limit of the circular loop, with $\textbf{c}$ a constant vector.
Consider a small perturbation along the transverse directions, $\delta \textbf{x}\, , \, \textbf{c}\cdot \delta \textbf{x}=0 $.
At first order in $\delta x_i/x_i$, we find
\begin{equation}
    \delta \ddot{x}_i+\lr{2\frac{\dot a}{a}+\frac{\dot \mu}{\mu}}\delta \dot{x}_i-\delta x_i''=0\, .
\end{equation}
This equation is solved by
\begin{equation}\label{eq:perturbative-solution}
    \delta x = \textbf{A}_k \lr{\frac{\tau_0}{\tau}}^{\nu} J_\nu (k\tau)e^{ik\zeta^1}+
    \textbf{B}_k \lr{\frac{\tau_0}{\tau}}^{\nu} Y_\nu (k\tau)e^{ik\zeta^1}\, ,
\end{equation}
where $\nu=(q-1)/2$, while $J_\nu (k\tau)$ and $Y_\nu (k\tau)$ are Bessel functions of the first and second kind, respectively\footnote{The second Bessel function only solves the equations for $m \, n >4$.}.
We will henceforth set $\textbf{B}_k=0$ noting that the conclusions below also apply for these solutions.
Some comments about the validity of the perturbative expansion are in order.
We have imposed $\delta x' \ll \partial_{\zeta^1} \zeta^1 =1$, which implies $k \delta x \ll 1$.
That is, we cannot resolve fluctuations of amplitude of the order of their wavelength or smaller.
In addition, we have $\partial_\tau \delta x \sim \lr{k-\nu/\tau}A_k \ll 1$, which is implied by the previous constraint for subhorizon fluctuations.
It is useful to use the large-argument expression for the Bessel function:

\begin{equation}\label{eq:subhorizon-bessel}
    \lim_{k\tau \to \infty} J_\nu (k\tau)=  \sqrt{\frac{2}{\pi k\tau}} \cos \lr{z-\frac{\nu \pi}{2}-\frac{\pi}{4}}\, ,
\end{equation}

where we note that the amplitude is $\nu$-independent.
In the subhorizon limit, the solutions are therefore oscillatory with a  growing/decreasing amplitude $\delta \textbf{x}\sim \tau^{q/2}\sim t^{m/2}/a(t)$.

We are now in a position to argue that the momentum parameter receives negative contributions from the small scale structure of the string whenever $mn>4$.
To do so, note that the (unnormalized) momentum parameter is given by

\begin{equation}
    \langle (1-\dot{\textbf{x}}^2) \dot{\textbf{x}} \cdot \textbf{u} \rangle = \langle \frac{R}{a}\frac{1}{\epsilon^2} \textbf{x}'' \cdot \dot{\textbf{x}} \rangle \, .
\end{equation}
For our solutions, $\textbf{x}''=-k^2\delta \textbf{x}$ and $\dot{\textbf{x}}=-\textbf{A}_k\lr{\frac{\tau_0}{\tau}}^\nu J_{\nu+1}(k\tau)e^{ik\xi^1}$.
The curvature radius is $R= 1/4k$, and $\epsilon=1+\mathcal{O}(\delta \textbf{x}^2)$.
The averaging procedure over $\zeta^1$ results in a factor of $1/2$.
More interesting is the time-average, which can be shown to produce
\begin{equation}
    \langle \frac{R}{a}\frac{1}{\epsilon^2} \textbf{x}'' \cdot \dot{\textbf{x}} \rangle=
    \frac{1}{16a}\textbf{A}_k^2k^2
    \lr{\lr{\frac{\tau_0}{\tau}}^{2\nu}J_\nu (k\tau)^2-\frac{1}{2^\nu \Gamma(1+\nu)^2}}\, ,
\end{equation}
where $\Gamma(x)$ is Euler's gamma function and we have averaged, for simplicity, from an initial time $\tau=0$.
The crucial point here is that for subhorizon perturbations, using Eq.~\eqref{eq:subhorizon-bessel}, the first term dominates for $2\nu+1<0$, which is equivalent to $mn<4$.
Otherwise the second term dominates and the contribution to the momentum parameter is negative - we take this as suggestive evidence of a flip of sign in the momentum parameter for the VOS model when $mn>4$ and therefore for the existence of scaling solutions in this regime, as explained below.
We believe this is so even if our solutions do not fully capture the complicated microscopic configuration of the long strings in a scaling solution.
In particular, the microscopic equations cannot take into account the loss of energy into loops by interactions of the strings, which is a key ingredient for scaling to occur.

\subsection{Long strings}\label{sec:long-strings}

In this section, we examine how the modified equations of motion \eqref{eq:E0vt}-\eqref{eq:Eivt} will impact the evolution and properties of the network, focusing on the population of the long strings.

\subsubsection{Network evolution}
From \eqref{eq:E0vt}, the energy density $\rho \sim \frac{E}{a^3}$ of strings then obeys the equation 
\begin{equation}\label{eq:rho1}
    \frac{{\rm d} \rho}{{\rm d} t} = - 2H(1+v^2) \rho +\frac{1}{\mu} \frac{{\rm d} \mu}{{\rm d} t}(1-v^2) \rho
\end{equation}
where we have switched to the FRW time coordinate $t$. Notice that, in the non-relativistic limit $(v \rightarrow 0)$, the energy density redshifts as $\rho \sim \mu(t) /a(t)^2$, while in the ultra-relativistic limit $(v \rightarrow 1)$ it behaves as radiation, $\rho \sim 1 /a(t)^4$. From the definition of the characteristic length $\rho = \frac{\mu}{L^2}$, one can turn \eqref{eq:rho1} into an equation for $L$. For long strings, one also has to take into account energy loss into loops, which can be modelled as
\begin{equation}\label{eq:rho2}
    \frac{{\rm d} \rho}{{\rm d} t}  =- \tilde{c} v \frac{\rho}{L}.
\end{equation}
Putting everything together, \eqref{eq:rho1}-\eqref{eq:rho2} can be combined to give
\begin{equation}\label{eq:L}
    2 \frac{{\rm d} L}{{\rm d}t}= 2 H L(1+v^2)+\tilde {c}v+\frac{L}{\mu} \frac{{\rm d} \mu}{{\rm d}t} v^2.
\end{equation}
Following the same steps, one can use \eqref{eq:Eivt} to derive a new equation for $v$, namely 
\begin{equation}\label{eq:dvdt}
   \frac{{\rm d} v}{{\rm d}t} = (1-v^2) \left[\frac{k(v)}{L}-\left(2 H + \frac{1}{\mu} \frac{{\rm d} \mu}{{\rm d}t} \right) v \right].
\end{equation}
The difference between these and those of~\cite{Martins:2018dqg} were explained in footnote~\ref{ft:eoms}.
Here, one can already see a crucial, qualitative difference with respect to the constant tension case: if $2 H + \frac{1}{\mu} \frac{{\rm d} \mu}{{\rm d}t} <0$, the RHS of \eqref{eq:dvdt} cannot be negative for any $v$ satisfying $k(v)>0$, forbidding the existence of a fixed point in the usual regime $v^2 \leq \frac{1}{2}$. Furthermore, if $m n  > 4$, the asymptotic velocity will satisfy $ v^2 > \frac{1}{2}$, regardless of whether there is a scaling regime or not, and we have provided evidence in Section \ref{sec:momentum-parameter} that $k(v)<0$ in this regime.
All of this suggests the existence of a fixed point with $v^2 > \frac{1}{2}$ in this case. To study the general case, it is useful to perform a full dynamical system analysis of the two evolution equations, as carried out in the next section (see Appendix \ref{app:ds} for more details). Before we continue, let us now make a brief comment on possible scaling solutions, \emph{i.e.} stable (attractor) solutions with constant $v = \bar{v}$ and $L/t = \bar{\xi}$. As already noticed in \cite{Yamaguchi:2005gp}, there is one crucial physical difference with respect to the constant tension case: the \emph{overall} energy density in cosmic strings now scales as
\begin{equation}\label{eq:rhomu}
    \rho = \frac{\mu}{L^2} = \frac{\mu_0 t_0^{m}}{\xi^2 t^{2+m}},
\end{equation}
and it will decrease with respect to the background if $m> 0$. In particular, scaling solutions will not be characterised by a constant $\Omega_s= \rho_s / 3 H^2$, as happens in the constant tension case.

\subsubsection{Dynamical system analysis}

The evolution equations \eqref{eq:L}-\eqref{eq:dvdt} can be recast as the autonomous system
\begin{equation}\label{eq:sys}
    \begin{cases}
        \frac{{\rm d} \xi}{{\rm d}z} = \xi \left[\left(\frac{2}{n}-1 \right)+\frac{v^2}{2}\left(\frac{4}{n}-m \right) \right]+ \frac{\tilde{c}(v)v}{2}\\
         \frac{{\rm d} v}{{\rm d}z} = (1-v^2) \left[\frac{k(v)}{\xi}-\left(\frac{4}{n} -m\right) v \right],\\
    \end{cases}
\end{equation}
through the change of variable $z = \log t$ and $L = \xi(t) t$. As in the previous sections, we assume a universe dominated by a background fluid with $\rho \sim a^{-n}$ and $\mu \sim t^{-m}$, with $n>2$ and $m>0$. Note that in the first equation we have not included the gravitational back-reaction, \emph{i.e.} the effect of GW production, which gave a sub-leading effect in the contracting universe analysis of \cite{Avelino:2002hx,Avelino:2002xy}. In our case, due to the decreasing tension, we expect that the effect will be negligible. Moreover, the attentive reader will have noticed that the chopping parameter $\tilde{c}$ has been promoted to a velocity-dependent function $\tilde{c}(v)$. Although so far we have assumed $\tilde{c}$ to be a constant, for $ mn > 4$ the network is expected to probe the ultra-relativistic regime, where relativistic corrections become important. Taking Lorentz contraction of the correlation length into account, a reasonable ansatz is $\tilde{c}(v) \simeq \tilde{c} (1-v^2)^{1/2}$ \cite{Avelino:2002hx,Avelino:2002xy}, which implies
$\tilde{c}\rightarrow 0$ in the ultra-relativistic limit. However, there may also be additional loss terms in the first equation (for example decays to other particles) that we have not taken into account, potentially leading to a small, effective $\tilde{c} \neq 0$. For this reason, following \cite{Avelino:2002hx,Avelino:2002xy}, we will also consider the case where $\tilde{c}(v)$ asymptotes to a constant $\tilde{c}_{\infty} \neq 0$ as $v \rightarrow 1$.

In this form, \eqref{eq:sys} can be analysed with standard techniques for dynamical systems. In particular, one can distinguish two different classes of fixed points, examined in detailed below. The first one is that of scaling fixed points, which straightforwardly generalise the scaling solution for $m=0$. The second class consists of ultra-relativistic fixed points, where $v$ asymptotes to $1$. These are reminiscent of what happens for a network of cosmic strings in a contracting universe \cite{Avelino:2002hx,Avelino:2002xy}, with a few differences that will be reviewed. Here we will just state the main outcome of the analysis, leaving the detailed proofs to Appendix \ref{app:ds}.

\paragraph{Scaling fixed point(s)}\mbox{}\\
One can easily verify the existence of a scaling solution with $L = \xi t$ and 
\begin{equation}\label{eq:fpsc}
  \xi= \sqrt{\frac{n}{8} \frac{k(\bar{v})\left(k(\bar{v}) + \tilde{c} \right)}{\left(1-\frac{m n}{4} \right) \left(1-\frac{2}{n} \right)}} \quad \quad \quad \quad \bar{v} = \sqrt{\frac{n}{2} \frac{k(\bar{v})}{k(\bar{v})+\tilde{c}} \frac{1-\frac{2}{n}}{1-\frac{mn}{4}}}, 
\end{equation} 
which reduce to \eqref{eq:vos1} and \eqref{eq:vos2} for $m=0$. The multiplicity of the solutions \eqref{eq:fpsc}
 depends on the possible values of the velocity $\bar{v}$, determined by the solutions to
\begin{equation}\label{eq:zeros}
    k(\bar{v}) - \frac{\tilde{c} \bar{v}^2}{\frac{2(n-2)}{4-mn}-\bar{v}^2} =0.
\end{equation}
For $mn < 4$ there is always one such fixed point, which has $\bar{v}^2 < \frac{1}{2}$ and is always an attractor. It corresponds to the usual fixed point found for $m=0$. For $ mn >4$, solutions to \eqref{eq:zeros} always come in pairs for a momentum parameter of the form \eqref{eq:k(v)}. However, only the one with lowest value of $v^2$ will be stable, while the other one is always a saddle. From \eqref{eq:fpsc}, it is also easy to see that if $mn > 4$, such solutions must have $k(\bar{v}) \leq 0$, while $k(\bar{v}) \geq 0$ if $ mn <4$. As we have just mentioned, this traces back to the fact that, if $ 2 H + \frac{1}{\mu} \frac{{\rm d} \mu}{{\rm d}t} <0$, the RHS of \eqref{eq:dvdt} will always be positive, and $v$ will keep growing until $k(v)$ becomes negative.\footnote{Notice that this is the same condition required for loops to grow faster than the scale factor, as shown in \cite{Conlon:2024uob}.} In particular, the phenomenological forms of $k(v)$ proposed in the literature (See Eq.\eqref{eq:k(v)}), which have been validated through numerical simulations, are all strictly positive for $v^2 \leq \frac{1}{2}$.
In section~\ref{sec:momentum-parameter}, we also provided analytic evidence that small scale structure induces a negative $k(v)$ whenever $mn>4$.

\paragraph{Ultra-relativistic fixed point, constant chopping parameter}\mbox{}\\
If $\tilde{c}(v) \rightarrow \tilde{c}_{\infty}$ as $v \rightarrow 1$, there is new, ultra-relativistic fixed point, located at\footnote{Since $v=1$ is inconsistent, such a fixed point can never be reached in finite physical time.}
\begin{equation}\label{eq:ulp}
    v_r= 1 \quad \quad \quad \xi_r = \frac{\tilde{c}_{\infty}n}{2(n-2)n+mn-4}.
\end{equation}
It is stable for
\begin{equation}
    mn > {\rm Max} \left\{ 4, 4+n \left(\frac{m}{2}-1 \right)\right\},
\end{equation}
so it can never be an attractor for $ mn <4$. The solution to \eqref{eq:ulp} is of the genuine scaling type, so its energy density will behave as in \eqref{eq:rhomu}. However, if $\tilde{c} \ll \xi_0$, as expected in practical applications, trajectories approaching the fixed point will effectively behave as if $\tilde{c}_{\infty} =0$ for a time 
\begin{equation}
    \frac{t_r}{t_0} \lesssim \left( \frac{\xi_0}{\tilde{c}_{\infty}} \frac{2(n-2)+mn-4}{n} \right)^{\frac{2n}{2(n-2)+mn-4}}.
\end{equation}
In this timeframe, the network behaves like a gas of very relativistic strings, with an equation of state equal to that of radiation and $\rho \sim a^{-4}$. This is analogous to the case of cosmic strings in a contracting universe \cite{Avelino:2002hx,Avelino:2002xy}. However, a crucial difference is that in this case it can coexist with a stable scaling fixed point, and the outcome of the system crucially depends on the initial conditions, as well as the values of the various parameters.

\paragraph{Ultra-relativistic fixed point, zero chopping parameter}\mbox{}\\ 
If $\tilde{c}(v)$ approaches $0$ in the ultra-relativistic regime, the results of the previous paragraph cannot be applied, as $\xi=0$ is a singular point of the system \eqref{eq:sys}. However, for a chopping parameter of the form $\tilde{c} (1-v^2)^{\alpha}$, it can be shown that there still exists a fixed point with $v =1$ and $\xi  \rightarrow 0$. For $\alpha = \frac{1}{2}$, as motivated above, the latter is an attractor if
\begin{equation}
    m > \frac{16+2n}{5n}.
\end{equation}
Furthermore, it is characterised by
\begin{equation}
     \xi(t) = \xi_0 \left(\frac{t}{t_0}\right)^{-\frac{1}{2n}\left[ 2(n-2)+ mn -4 \right]} \quad \quad \text{and} \quad \quad \rho= \frac{\mu(t)}{L(t)^2} \sim \frac{1}{t^{8/n}} \sim \frac{1}{a(t)^4},
\end{equation}
so the network effectively behaves as radiation. For these reasons, we will also refer to this as the radiation fixed point. From  numerical solutions to the equations of motion, one can also infer that, similarly to the previous case, even when the fixed point is a saddle there can be trajectories approaching the attractor which still contain long epochs where the network closely resembles radiation, even if the endpoint of the evolution is a different fixed point.

\subsubsection{Dependence on initial conditions}

We have shown that for a vast range parameter space two qualitatively different and stable fixed points can coexist, corresponding to a proper scaling solution and a radiation fluid. Moreover, even when only the scaling fixed point is an attractor in the true sense of the word, the radiation-like solution can still behave like a ``fake" attractor, and lead to long, late-time epochs where the network redshifts in the same way as radiation. The latter, in particular, is true both in the cases where $\tilde{c}_{\infty} >0$, and $\tilde{c} (v) \rightarrow 0$. 

\begin{figure}
\centering
\begin{subfigure}[b]{0.45\textwidth}
    \centering
  \includegraphics[width=\textwidth]{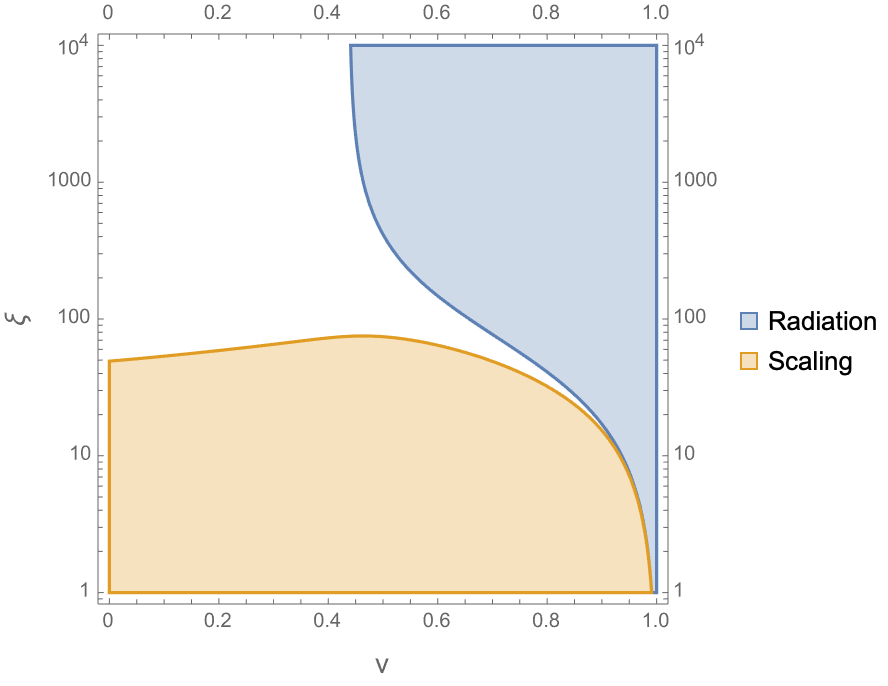}
  \caption{$z=10; \, m=1,n=6$}
  \label{fig:1a}
\end{subfigure}
\begin{subfigure}[b]{0.45\textwidth}
  \centering
  \includegraphics[width=\textwidth]{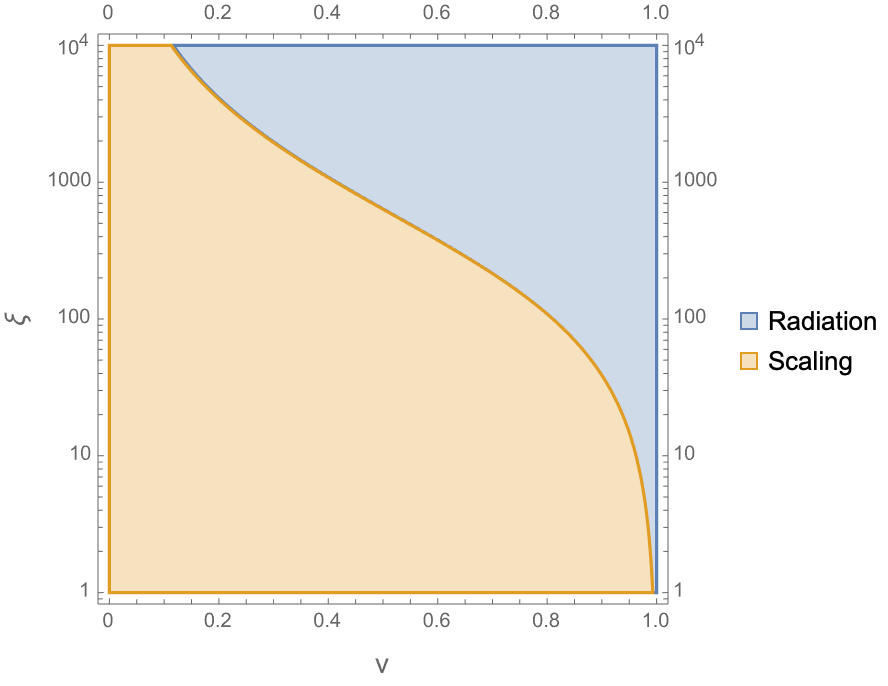}
  \caption{$z=20; \, m=1,n=6$}
  \label{fig:1b}
\end{subfigure}
\caption{Endpoint of the evolution as a function of the initial conditions. The blue and orange regions denote points that are approaching (see text) the scaling and radiation-like attractors respectively after having evolved for a time $\log t= z=10$ and $\log t= z=20$. The white region depicts points that still are far enough from both attractors after the same amount of time.}
\label{fig:1}
\end{figure}

Therefore, the evolution of the system will be heavily dependent on initial conditions, and the timescales one is interested in. In this last section, we would like to give a few examples for realistic values of the parameters. For the rest of the section, we will further assume $\tilde{c}(v) = \tilde{c} (1-v^2)^{1/2}$, with $\tilde{c}=0.23$ \cite{Martins:2000cs}. We can start from the phenomenologically interesting case $m=1,n=6$, which corresponds to volume kination in a string-theoretic embedding. In Figures \ref{fig:1a}-\ref{fig:1b}, we plotted regions where the late-time value of $v$ is close to either critical points after a time $t=e^{10}$ and $t=e^{20}$,\footnote{By convention, we have arbitrarily defined this as $|v-\bar{v}|< 0.01$.} as a function of initial conditions for $\xi$ and $v$. The difference between the two plots shows how it may take a relatively long time to reach one of the attractors, and even how trajectories that stay very close to one of the fixed point for a long time can still end up converging to the other one. At network formation, the density of strings per Hubble patch is of order $ \mathcal{O}(1)$, and we expect $\xi \sim \mathcal{O}(1)$. Therefore, the proper scaling fixed point will most likely be reached at very late times, unless the initial velocity is very close to $1$. 

\begin{figure}[h!]
\centering
\begin{subfigure}[b]{0.45\textwidth}
    \centering
  \includegraphics[width=\textwidth]{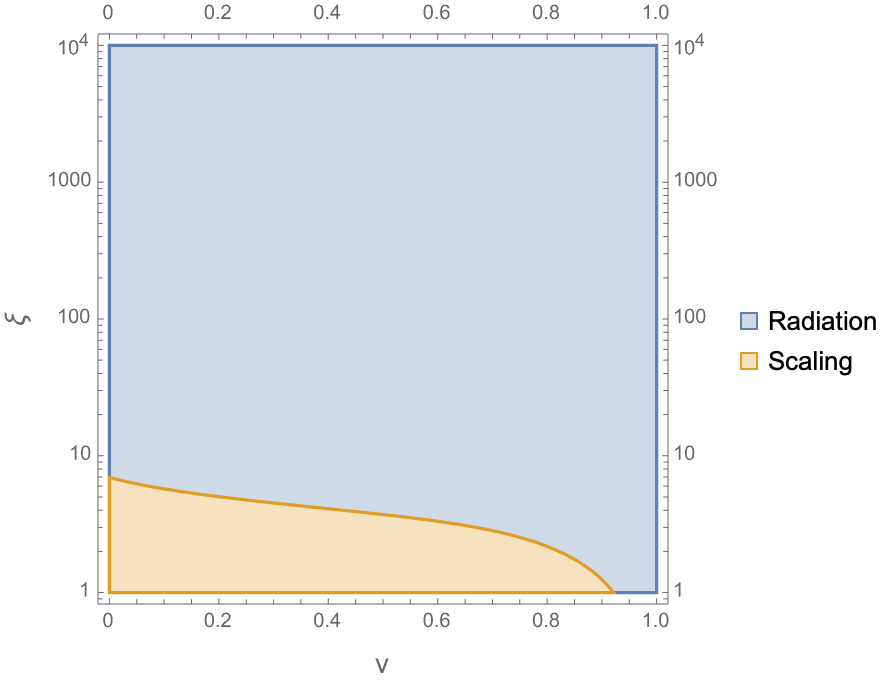}
  \caption{$z=20;\, m=2,n=3$}
  \label{fig:2a}
\end{subfigure}
\begin{subfigure}[b]{0.45\textwidth}
  \centering
  \includegraphics[width=\textwidth]{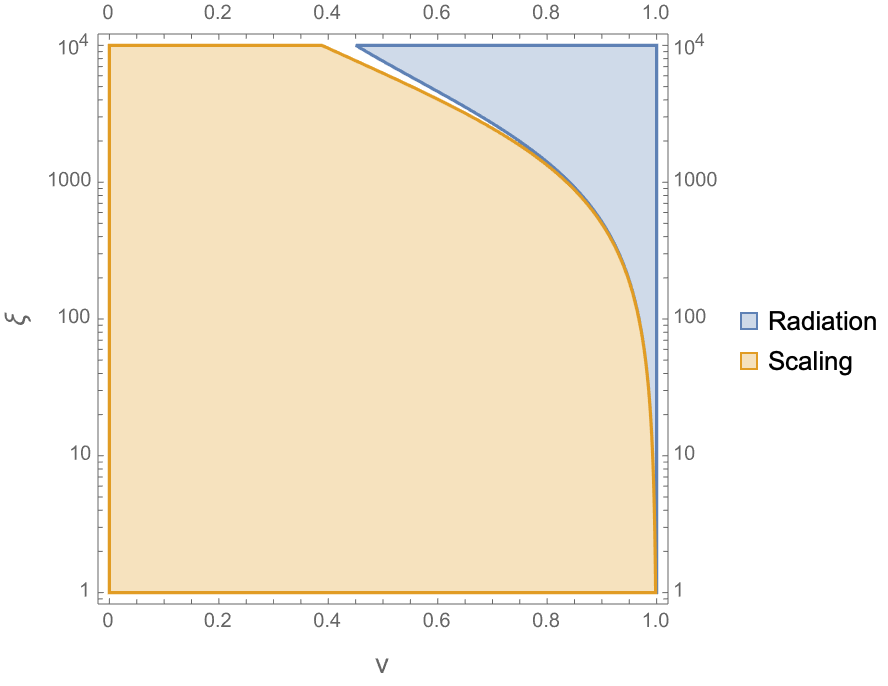}
  \caption{$z=20;\, m=1,n=5$}
  \label{fig:2b}
\end{subfigure}
\caption{Endpoint of the evolution as a function of the initial conditions. The blue and orange regions denote points that are approaching (see text) the scaling and radiation-like attractors respectively after having evolved for a time $\log t= z=20$. }
\label{fig:2}
\end{figure}

However, this result is very specific to these values of $m$ and $n$. In Figures \ref{fig:2a}-\ref{fig:2b}, we show the same plot as in \ref{fig:1a}-\ref{fig:1b}, for different values of $m$ and $n$, keeping everything else unchanged. Small variations of either one can drastically alter the picture above, leading for example to situations where the radiation-like fixed point is a much more likely future outcome.

We end with a note of caution: the VOS model for a time-dependent tension has not been validated through numerical simulations, and is therefore on a less solid footing than the constant tension case. Moreover, the numerical values for some of the phenomenological parameters used to describe the network have to be extracted from simulations, which usually assume a constant tension and have only been performed for standard backgrounds such as radiation or matter domination (and transients between them). Since $\tilde{c}$ is expected to be constant regardless of the background regime, we will make use of this estimate for different epochs (such as kination), and also extrapolate it to the case of a time-dependent tension.
It would nevertheless be interesting to test these assumptions with an explicit numerical analysis, which we defer to future work.

\subsection{Loop evolution}\label{ssc:loops}
Having concluded that a scaling regime is achieved when the tension varies, we now study the averaged dynamics of subhorizon loops.
For a single loop, one can similarly integrate \eqref{eq:E0vt} to obtain the evolution of its total energy $E_{\ell}\equiv \mu \ell$:
\begin{equation}
    \frac{{\rm d} E_{\ell}}{{\rm d}t} = E_{\ell} H(1-2v_{\ell}^2) + E_{\ell} \frac{1}{\mu}\frac{d\mu}{dt} (1-v^2)- \Gamma' G \mu^2 v_{\ell}^6,
\end{equation}
where the energy loss due to emission of gravitational radiation has been added in by hand, and $\Gamma' = 8 \Gamma$. Since loops are typically smaller than the horizon size, one can approximate their velocity as $v_{\ell}^2 =\frac{1}{2}$, which holds exactly in flat space.
Then, the first term cancels out and one obtains 
\begin{equation}
    \frac{{\rm d} E_{\ell}}{{\rm d}t} =  E_l \frac{1}{2\mu}\frac{d\mu}{dt} - \Gamma G \mu^2,
\end{equation}
which matches with Eq. (34) of \cite{Emond:2021vts} for $\mu \sim \frac{1}{a^2}$. For a decreasing tension, the energy of the loops will also decrease with time as well. It also follows that the time evolution of the loop length is given by
\begin{equation}\label{eq:lt}
   \frac{{\rm d} \ell}{{\rm d}t} =  - \ell \frac{1}{2\mu}\frac{d\mu}{dt}- \Gamma G \mu,
\end{equation}
where one of the terms has (crucially) changed sign, now allowing for the loop to grow even if it is emitting gravitational radiation. Since $\ell \lesssim t^{m/2}$, for $m <2 $ the loops will always grow slower than the horizon size. This justifies a posteriori the assumptions that $v_{\ell}^2 = \frac{1}{2}$, true on sub-horizon scales. Notice that if the tension is decreasing with time, the first contribution to \eqref{eq:lt} is positive, and if it dominates over energy loss loops can in principle \emph{grow} with time.  With the specific dependence $\mu (t) = \mu_0 \left(\frac{t_0}{t}\right)^m$, this is solved by

\begin{equation}\label{eq:loop-ev-vary-tension}
    \ell(t)=\lr{\frac{t}{t_i}}^{m/2} \lr{\alpha-\frac{\Gamma G \mu_0}{\frac{3m}{2}-1}\lr{\frac{t_0}{t_i}}^m}t_i+\frac{\Gamma G\mu_0}{\frac{3m}{2}-1}\lr{\frac{t_0}{t}}^{m-1}t_0 \,
    ,
\end{equation}
for $m \neq 2/3$, and
\begin{equation}\label{eq:loop-tracker}
    \ell(t)=\left[\alpha  - \Gamma G\mu_0 \lr{\frac{t_0}{t_i}}^{2/3}\log \lr{\frac{t}{t_i}}\right]\lr{\frac{t}{t_i}}^{1/3}t_i \, .
\end{equation}
for $m=2/3$.
Here we have considered that the loop is formed at time $t_i$, and a tension $G\mu (t)$ which satisfies $G\mu (t_0)= G \mu_0 \ll \alpha$ (the inequality assumes a large hierarchy between the string and Planck scales, which is required for control over the calculations).
Therefore, we find that if $0<m \leq \frac{2}{3}$ the physical length of the loops can temporarily grow at the beginning of their evolution, but they will still shrink in size and approach zero asymptotically.
If $m> \frac{2}{3}$, however, they will grow monotonically.

\subsection{Percolation}

As a consequence of the decreasing tension, it was noticed in the recent paper \cite{Conlon:2024uob} how loops on sub-horizon lengths will grow in size more rapidly than the scale factor if $2\mathcal{H}+\frac{\dot{\mu}}{\mu}<0$ is satisfied. If their density is large enough, such loops can merge with each other to form larger ones and eventually percolate, giving rise to a network of long strings. While such a mechanism was introduced in~\cite{Conlon:2024uob} to provide an alternative scenario for the formation of a cosmic (super)string network, one might ask if percolation itself may not make a pre-existing string network unstable. Indeed, as we have seen in the previous sections the equilibrium (scaling) configuration of any cosmic string network contains a population of loops arising from the reconnection of long strings. If such loops were able to percolate, and convert their energy back into a network of long strings, one might be worried that they would alter the properties of the network, and perhaps even overclose the universe.

For this reason, let us estimate how long it would take for percolation to occur, assuming the existence of a network from the time $t_0$. For a fixed final time $t$, the length distribution of the loop population can be derived by the production rate \eqref{eq:lprod}, as well as the fact that 
a loop produced at the time $t_p$ will have grown as
\begin{equation}
    \ell(t) = \alpha t_p \left( \frac{t}{t_p}\right)^{m/2},
\end{equation}
where we are ignoring subleading terms due to GW emission.
The final expression is
\begin{equation}\label{eq:loopd}
    \frac{{\rm d} n}{{\rm d} \ell} =  \frac{2\mathcal{F}_\alpha \alpha^{\frac{6}{2-m}}}{(2-m)} \frac{c \bar{v}}{\gamma \xi^{3}}    \frac{t^{\frac{3m}{2-m}} }{\ell^{\frac{8-m}{2-m}}}.
\end{equation}
As in \cite{Conlon:2024uob}, percolation may be estimated to have occured when the typical string separation is the same order as the string length, namely
\begin{equation}
    \int_{\ell(t_0)}^{\ell(t)} {{\rm d} \ell}\,  \frac{\ell(t)^3}{a(t)^3} \frac{{\rm d} n}{{\rm d} \ell} \sim \mathcal{O}(1).
\end{equation}
The main contribution from the integral comes from loops produced at the time of network formation $t_0$ (the integrand may well be approximated by a delta function), resulting in the condition
\begin{equation}
    \frac{t}{t_0} \lesssim \left[ \frac{3m}{2\mathcal{F}_\alpha} \frac{\gamma \xi^{3}}{c \bar{v}} \alpha^{\frac{3m-6}{2-m}} \right]^{ \frac{2n}{3(mn-4)}}.
\end{equation}

Therefore, the duration of time for which the scaling regime described in Section \ref{sec:long-strings} can survive is sensitive to the microscopic parameters describing the network. 
It is worth noting that in the case of superstrings these will inherit a dependence in the intercommutation probability and so the timescales are to be evaluated on a model-by-model basis.
A more systematic treatment is left for future work.
In the case of volume kination $m=1,n=6$ and substituting typical numerical values for the parameters
\begin{equation}
    \frac{t}{t_0} \lesssim 10^6-10^7.
\end{equation}
This is a sizeable fraction (on a log scale) of the maximal allowed lenght of kination in LVS, $\frac{t}{t_0} \sim 10^{10}$ \cite{Conlon:2024uob}.

The question of what happens afterwards is slightly more subtle, and we suspect it cannot be answered with the tools at our disposal.\footnote{Progress on this matter could be achieved using the Boltzmann equation approach of~\cite{Copeland:1998na,Frey:2023khe}.} Therefore, this paragraph should be considered mainly as speculative. As the loops merge with one another and start to percolate, they start to feed back into the population of long strings. To model this within the VOS framework, a cheap attempt may be to decrease the loop chopping efficiency $\tilde{c}$ governing the decays of long strings into loops. If $\tilde{c}$ remains lower bounded, then the network will also be exhibiting a quasi-scaling regime as per Eq. \eqref{eq:quasisc}. On the other hand, $L(t)$ may grow sub-linearly if $\tilde{c} \rightarrow 0$. Even if this is the case, the network can never dominate the energy density of the universe, unlike what would happen  with a constant tension. Indeed, by \eqref{eq:quasisc}, 
\begin{equation}
    \rho=\frac{\mu}{L^2} \leq \frac{C}{t^2},
\end{equation}
for an appropriate constant $C$.

\section{Outlook}\label{sec:outlook}

In this paper, we analysed the dynamics of a cosmic string network with a time-dependent tension. This is motivated by the idea that, in String Theory constructions, the tensions of both fundamental and composite strings can depend non-trivially on the expectation values of the moduli. In turn, the latter can evolve as a function of time, and induce variations in the string tension. This could be very relevant for the very early stages of the universe's evolution, which are mostly unconstrained by observations until the onset of Big Bang Nucleosynthesis (BBN) \cite{Allahverdi:2020bys} and where string theoretic considerations motivate rolling scalars \cite{Cicoli:2024bwq}. As an example, we discussed the case where a time-dependent volume modulus in the very early universe results in decreasing string scale $m_s$ \cite{Apers:2024ffe}, potentially offering a way to mitigate observational constraints on the existence of cosmic superstrings \cite{Conlon:2022pnx}.

To this aim, we extended the Velocity One Scale (VOS) model of \cite{Martins:1996jp,Martins:2000cs} to describe a string network in the time-dependent case. This was achieved by considering the effect of new terms appearing in the microscopic equations of motion, and through a careful averaging on larger scales. In particular, we provided both microscopic and macroscopic arguments to suggest that the ansatz for the momentum parameter $k(v)$ proposed in \cite{Martins:2000cs} should also hold in the macroscopic case. The outcome of this analysis was presented in Section \ref{sec:long-strings}, and we also summarize it here.
We found that the well-known (scaling) attractor solution, with a linearly growing correlation length ($L = \xi t$), persists to the case of a varying string tension. Notice, however, that due to a decreasing tension the energy density in the network is no longer a fixed fraction of the background, and the scaling parameters are modified in accordance to Eqs \eqref{eq:fpsc}-\eqref{eq:zeros}. More surprisingly, we also noticed that if 
\begin{equation}\label{eq:muH}
     2 \mathcal{H}+ \frac{\dot{\mu}}{\mu} <0,
\end{equation}
a new attractor solution can exist, corresponding to a network that redshifts like radiation. In that case, the end state of the system depends on the initial conditions, although the standard scaling solution is preferred for realistic values of the latter. In Section \ref{ssc:loops}, we also discussed the evolution of the loop population taking gravitational radiation into account, explicitly confirming the finding of \cite{Conlon:2024uob} on the bizarre phenomena of loop growth when \eqref{eq:muH} is violated.

Our analysis could be extended in several directions, and some points are still unanswered. Here, we list a few promising directions for future explorations:

\begin{itemize}
    \item The VOS model and some of its extensions (such as superconducting strings) have been numerically verified in a number of cases, see for example \cite{Moore:2001px,Hindmarsh:2017qff,Correia:2019bdl,Martins:2020jbq,Auclair:2022ylu}. To put our results on firmer grounds, it would be necessary to carry out numerical simulations of cosmic strings with time-varying tensions. This could be used to both test some of our key assumptions (such as the functional form of $k(v)$) and also the final results on the attractor solutions.
    \item Our microscopic, analytic understanding of the momentum parameter $k(v)$ is only limited to certain particular configurations, such as the static, infinite string. Is it possible to find a general argument showing that $k(v)$ switches sign when $v^2 = \frac{1}{2}$, in accordance with the ansatz \eqref{eq:k(v)}?
    
    \item It would be interesting to examine more in depth the effect of a varying tension for realistic axion string networks (see \cite{Benabou:2023npn} for a recent analysis on the cosmology of string theory axion strings).\footnote{For completeness, we should mention that a logarithmic violation of the scaling regime has been observed  for axion strings \cite{Gorghetto:2018myk,Gorghetto:2020qws}, due to the slowly-varying, IR contribution to the tension.} Are there any natural string theory realisations of this scenario, and would they lead to any observable signatures?
    
\end{itemize}

Moreover, our results open the door to a computation of the gravitational wave spectra originating from time-dependent string networks, which we will report on in the future \cite{us:2024?}. 
On a qualitative level, we expect this scenario to release a larger signal at high frequencies, when compared to the constant tension case.
The reasoning is as follows: fixing a late-time tension $G\mu_0$, the low-frequency spectra of our scenario and a standard scenario with constant tension $G\mu_0$ should be indistinguishable, since low frequencies are sensitive to late times.
However, a key feature of GWs from cosmic strings is that they can carry information about the full cosmic history, including the early times at high frequencies.
At high frequencies, thus, it is natural to expect a growth of the signal as compared with the constant tension case, since GW spectra arising from networks of higher tension are larger.
The spectrum, however, depends on the evolution of the loop distribution, which is qualitatively different when the tension varies, so further work is required.

As already remarked in the introduction, the GW spectrum can only be reliably computed when the evolution of the network is well understood, so our results provide a first step in this direction. If a network of cosmic strings underwent a period in the early Universe on which the tension decreased, as suggested in scenarios featuring volume-modulus kination~\cite{Conlon:2022pnx,Conlon:2024ene,Conlon:2024hgw,Conlon:2024uob} or other displacements of the volume modulus after inflation, as in~\cite{Burgess:2022nbx,Cicoli:2024bwq}, we expect a modification in the GW spectrum at high frequencies, see~\cite{Aggarwal:2020olq} for a discussion of possible sources and detectors.\footnote{Another GW spectrum arising from long fundamental strings, this time in thermal equilibrium (which is possible if the strings feature additional decay channels which prevent the scaling regime) has been studied in~\cite{Frey:2024jqy}.} Modifications in the tilt of the spectrum due to a varying equation of state of the background have been studied before~\cite{Cui:2017ufi,Cui:2018rwi,Gouttenoire:2021jhk,Servant:2023tua,Blasi:2020wpy} and it would be interesting to compare these predictions with those arising from a period of varying tension.

\acknowledgments
We are grateful to Anish Ghoshal for participating in the initial stages of this project, as well as collaboration on the related work \cite{us:2024?}. We would also like to thank Joe Conlon, Ed Copeland and Noelia Sánchez González for conversations on their recent work \cite{Conlon:2024uob} and Amelia Drew, Joe Conlon, Fernando Quevedo, Paul Shellard and the anonymous referee for helpful comments on the manuscript and extensive discussions. We also thank the organisers of SUSY 23 and of the workshop WISPs in String Cosmology, where some of this work took place. The research of FR was partially supported by the Dutch Research Council (NWO) via a Start-Up grant and a Vici grant, and by a junior postdoctoral fellowship of the Fonds Wetenschappelijk Onderzoek (FWO), project number 12A1Q25N. The work of GV has been partially supported by STFC consolidated grant ST/P000681/1, ST/T000694/1. GV thanks the Institute for Theoretical Physics of Utrecht University for hospitality. This work  is partly based upon work from the COST Action COSMIC WISPers CA21106, supported by COST (European Cooperation in Science and Technology).

\newpage
\appendix

\section{Dimensional reduction: a primer}\label{sec:dim-red}
One of the main motivations for this article is the fact that the four-dimensional Planck scale is determined dynamically in string theory.
In this Appendix, aiming for self-consistency, we review the well-known process of dimensional reduction (see e.g.~\cite{Rattazzi:2003ea}).
Let us consider the simple example of the dimensional reduction of a 5-dimensional massless scalar theory on a circle:
\begin{equation}
S_5=\int{d^4x \, dy \lr{-\partial_M \phi^* \partial^M\phi}}\, ,
\end{equation}
where $M \in \lbrace{0,1,2,3,4\rbrace}$ and $ y \cong y+2\pi R$.
The field admits a Fourier expansion of the form
\begin{equation}
\phi(x,y)=\sum_n \phi^{(n)} (x) e^{i n y/R}\, ,
\end{equation}
and the equations of motion read
\begin{equation}
\lr{\partial_\mu \partial^\mu +\lr{\frac{ n}{R}}^2}\phi^{(n)}(x)=0\, 
\end{equation}
with $\mu \in \lbrace 0,1,2,3\rbrace $.
These are the equations of motion of an infinite family of four-dimensional massive scalar fields of masses $n/R$.
One can similarly integrate the action along the circle to find an effective 4D action
\begin{equation}
S_4=2\pi R \int{d^4x\, \sum_n \lr{-\partial_\mu \phi^{*\, (n)}\partial^\mu \phi^{(n)}- \lr{\frac{n}{R}}^2 |\phi^{(n)}|^2}}\, .
\end{equation}
Which again describes the four-dimensional physics of an infinite set of massive scalar fields.
The key insight arises at energies lower than the \textit{Kaluza-Klein} scale, $1/R$.
All fields except the zero mode can be integrated out and the remaining dynamics is that of one scalar field.

Let us now add gravity.
Consider the 5-dimensional Einstein-Hilbert action:
\begin{equation}
S_5=\frac{M_{p,5}^3}{2}
    \int{d^4 x\, dy
    \sqrt{-G} R_{(5)}}\, .
\end{equation}
It is easy to show that the equations of motion admit solutions of the form $\mathbb{R}^{1,3}\times S^1$, where the size of the circle $S^1$ remains undetermined.
Similarly, the metric admits a Fourier decomposition and the action can be integrated upon the circle.
Let us write a metric of the form
\begin{equation}
G_{\mu \nu}=\lr{\frac{\sigma_0}{\sigma}}^{1/3}\lr{g_{\mu \nu}-\kappa^2 A_\mu A_\nu}\, , 
\qquad G_{\mu 5}=-\kappa \sigma^{2/3} A_\mu \lr{\frac{\sqrt{\alpha'}}{R}}^2\, , 
\qquad  G_{55}=\sigma^{2/3} \lr{\frac{\sqrt{\alpha'}}{R}}^2\, .
\end{equation}
The powers of $\sigma$ are judiciously chosen so that the dimensionally reduced low-energy theory does not feature kinetic mixing between $\sigma$ and $g_{\mu \nu}$, i.e: we choose to work in Einstein frame, and note $\sigma^{1/3}$ determines the volume of the internal space in units of $2\pi\sqrt{\alpha'}$.
As above, the different elements admit a Fourier decomposition and the action can be integrated along the circle.
Focusing on the zero-modes, we find
\begin{equation}
S_{EH}=\frac{ 2\pi \sqrt{\alpha'}\sigma_0^{1/3}\, M_{p,5}^3}{2}\int{d^4x\, \sqrt{-g}\lr{R_{(4)}-\frac{1}{4}\sigma F_{\mu \nu}F^{\mu \nu}-
\frac{1}{6}\frac{\partial_\mu \sigma\, \partial^\mu \sigma}{\sigma^2}}}+\dots\, ,
\end{equation}
Upon canonical normalization, this is a massless scalar coupled to four-dimensional gravity and a vector field, with coupling determined by the VEV of $\sigma$.
The size of the internal space fixes the four-dimensional Planck scale in terms of the higher-dimensional Planck scale, $M_p^2=2\pi \sqrt{\alpha'}\sigma_0^{1/3} M_{p,5}^3$.
In string theory, a 10D theory is similarly dimensionally reduced, with the higher-dimensional Planck scale known in terms of the string scale.
Thus, the 4D Planck scale is a derived quantity in string theory.

Lastly, it is worth noting that this choice of metric frame alters dimensionful quantities in e.g. a lower-dimensional brane, which inherit a dependence in the extra-dimensional size due to a coupling with $\sigma$.
To see this, notice the presence of the volume form $\sqrt{-h}$ in the DBI action, with $h=\text{det}(h_{ab}) $ and $h_{ab}$ the pullback of the metric on the worldvolume of the object.
This leads to a change in energies $E \to E (\sigma_0/\sigma)^{1/6}$.

\section{Semi-analytic solution for the circular loop}\label{sec:sol-loop}
To make progress in the case where the expansion of the Universe is considered, we may use instead the fact that $\int{\varepsilon\, d\sigma}$ is the proper length of the loop, which grows if $\varepsilon$ grows.
Indeed, as pointed out in~\cite{Conlon:2024uob}, $\varepsilon$ measures the maximum invariant radius of the loop.
We have:
\begin{equation}
\varepsilon^2=\frac{R^2}{1-(\dot{R}L)^2}\, ,
\end{equation}
so that $\varepsilon$ coincides with $R$ when the velocity of $R$ is zero, i.e: at its maximum amplitude, and otherwise performs a Lorentz transformation from the center of mass frame of the string so that the proper length is invariant.
We thus conclude that varying $\varepsilon$ indicates, in general, that the proper length of the string $L$ changes.
For the circular loop, this implies that the maximum value of the function $R(t)$ grows or decreases, on average, and via Eq.~\eqref{eq:E0vt} it is clear that this depends in the sign of $2H+\frac{1}{\mu} \frac{{\rm d}\mu}{{\rm d}t}$.

We can be more concrete and find an analytic expression for the behavior of loops in an arbitrary power-law cosmology.
To do so, let us consider the formal solution to Eq.~\eqref{eq:E0vt}:
\begin{equation}\label{eq:eps-sol}
\varepsilon (t)=\varepsilon (0)\, e^{-q\int{\frac{d\tau}{\tau}\dot{\textbf{x}}^2}}\,\end{equation}
where $q=(4/n-p)/(1-2/n)$ for a power-law cosmology $a\sim t^{2/n}$ and varying tension $\mu \sim t^{-p}$.
We note that, due to causality, $\dot{\textbf{x}}^2\in [0,1]$, so at finite time this expression for $\varepsilon (t)$ has no zeros or poles.
This is to be contrasted with the definition of $\varepsilon$:
\begin{equation}
\varepsilon (\tau)=\frac{R(\tau)}{\sqrt{1-(\dot{R}L)^2}}\, 
\end{equation}
which, unless $R(\tau)$ and $1-(\dot{R}L)^2$ have the same zeros, has poles whenever the string gets relativistic $\dot{R}L=1$.
This suggests that we look for solutions satisfying
\begin{equation}\label{eq:v-ansatz}
\dot{R}(\tau)L=-\sin (f(\tau/L))\, . 
\end{equation}
This implies, via Eq.~\eqref{eq:eps-sol}, that the solution for $R(\tau)$ is given implicitly by
\begin{equation}\label{eq:r-sol-exp-app}
R(\tau)=e^{-q\int{\frac{d\tau}{\tau}\sin (f(\tau/L))^2}}\cos (f(\tau/L))\, ,
\end{equation}
for a function $f(\tau/L)$ which satisfies certain conditions described later.
It is worth noting, however, that $R(\tau)$ satisfies two properties, independently of the particular details of $f(\tau/L)$:
\begin{itemize}
\item $R(\tau)$ is an oscillating function whose frequency depends on the solution $f(\tau/L)$.
\item The maximum amplitude of $R(\tau)$ behaves as
\begin{equation}
R_{max}(\tau) \sim \lr{\frac{\tau}{\tau_0}}^{-q/2}\, ,
\end{equation}
together with a proportionality constant that depends  on the initial conditions and an oscillating function that relaxes to one.
In the case with constant tension, the amplitude radius therefore grows as $R(\tau)\sim 1/a(t)$, namely the physical radius remains constant, and as the tension varies we have $R(t) a(t) \sim t^{p/2}$.
\end{itemize}
We thus find that, regardless of the particular form of $f(\tau/L)$, whether the overall amplitude of the loops increases only depends in the sign of $q$, as pointed out in~\cite{Conlon:2024uob}.

Let us now go further and investigate the behaviour of $f(\tau/L)$.
Consistency of equations~\eqref{eq:v-ansatz} and~\eqref{eq:r-sol-exp-app} requires that $f(\tau/L)$ satisfies
\begin{equation}
e^{q\int{\frac{d\tau}{\tau}\lr{\sin \lr{f(\tau/L)}}^2}}=f'(\tau/L)+q\frac{L}{\tau}\cos (f(\tau/L)) \sin (f(\tau/L))\, .
\end{equation}
Thus, one may also write
\begin{equation}
R(t)=\frac{\cos (f(\tau/L))}{f'(\tau/L)+q\frac{L}{\tau}\cos (f(\tau/L)) \sin (f(\tau/L))}\, ,
\end{equation}
which leads to the following differential equation for $f(x)$:
\begin{equation}
2x^2 f''(x)-q x(1-3\cos (2f(x)))-q(1+q\sin (f(x))^2)\sin (2f(x))=0\, .
\end{equation}
We have not found simple solutions for this equation.
However, we can take limits $f(x)\ll 1$ and $f(x)\gg 1$.
In the former, we can linearize and write
\begin{equation}
x^2 f''(x)+q\, x f'(x)-q\, f(x)=0\, ,
\end{equation}
which admits the straightforward solution
\begin{equation}
f(x)=A_1x+B_1x^{-q}\, 
\end{equation}
for $q \neq -1$, and
\begin{equation}
    f(x)=A_1x+B_1x\log (x)\, 
\end{equation}
for $q=-1$.

In the case where $f(x)\gg 1$, we consider the averaged equation, finding 
\begin{equation}
2x^2 f''(x)-qxf'(x)=0\, ,
\end{equation}
which is solved by 
\begin{equation}
f(t)=A_2+B_2t^{1+q/2}\, ,
\end{equation}
for $q\neq -2$, and
\begin{equation}
f(t)=A+B\log (t)\, 
\end{equation}
for $q=-2$.

We thus find that, generically, $f(\tau/L)$ interpolates between $\tau^{q}$ and $\tau^{1+q/2}$.
That is, between $f(t)\sim a^2/t^p$ in the superhorizon limit to $f(t)\sim t^{1-p/2}$ in the subhorizon limit, recovering the Minkowskian limit $f(t)\sim t$ in the case with constant tension.
In our case of interest, where the initial conditions set $f(t_0/L)=0$ and $f'(t_0/L)=1$, $f(t_0/L)$ grows monotonically.

\section{Details on the dynamical system}\label{app:ds}
We would like to examine the structure of the autonomous, dynamical system \eqref{eq:sys}, which we rewrite here for ease of presentation:
\begin{equation}\label{eq:sysapp}
    \begin{cases}
        \frac{{\rm d} \xi}{{\rm d}z} = \xi \left[\left(\frac{2}{n}-1 \right)+\frac{v^2}{2}\left(\frac{4}{n}-m \right) \right]+ \frac{\tilde{c}(v)v}{2}\\
         \frac{{\rm d} v}{{\rm d}z} = (1-v^2) \left[\frac{k(v)}{\xi}-\left(\frac{4}{n} -m\right) v \right],\\
    \end{cases}
\end{equation}
We will consider two forms for the function $\tilde{c}(v)$, that is
\begin{equation}
    \tilde{c}(v)= \tilde{c}_{\infty} \quad \quad \text{or} \quad \quad \tilde{c}(v)= \tilde{c} (1-v^2)^{\alpha}, \quad 0 < \alpha < 1.
\end{equation}

\subsection*{Scaling fixed points}
Away from the ultra-relativistic limit, we can safely assume $\tilde{c}$ to be constant. Any fixed point with $v \neq 1$ must satisfy
\begin{equation}\label{eq:scfpA}
    \xi_{s}= \sqrt{\frac{n}{8} \frac{k(v_{s})\left(k(v_{s}) + \tilde{c} \right)}{\left(1-\frac{m n}{4} \right) \left(1-\frac{2}{n} \right)}} \, , \quad \quad \quad \quad v_{s} = \sqrt{\frac{n}{2} \frac{k(v_{s})}{k(v_{s})+\tilde{c}} \frac{1-\frac{2}{n}}{1-\frac{mn}{4}}},
\end{equation}
which formally corresponds to the usual VOS model attractor. Notice, however, that these are implicit equations, and the existence and number of solutions cruciallly depends on the form of $k(v)$. Assuming the usual ansatz \eqref{eq:k(v)}, this already tells us that $v_{s}^2 < 1/2$ if $m n <4$, and $1/2 <v_{s}^2 < 1$ if $mn >4$. More precisely, the possible values of $v_{s}$ can be determined from the zeroes of the function
\begin{equation}\label{eq:f(v)}
    \mathcal{F}(v)=k(v) - \frac{\tilde{c} v^2}{\frac{2(n-2)}{4-mn}-v^2}.
\end{equation}
Their stability depends on the eigenvalues of the Jacobian
\begin{equation}\label{eq:jac}
   J_s=
    \begin{pmatrix}
    -\frac{\tilde{c} v_s}{2 \xi_s }    & k(v_s)+\frac{\tilde{c}}{2}  \\
-\frac{(1-v_s^2)k(v_s)}{\xi_s^2}   \,\,\,\,\,\,      &           (1-v_s^2) \left[ \frac{k'(v_s)}{\xi_s}-\left(\frac{4}{n}-m\right)\right]   \\
    \end{pmatrix},
\end{equation}
where all quantities are understood to be evaluated at the fixed point. For later reference, the trace and determinant of the Jacobian at the point(s) \eqref{eq:scfpA} can be expressed as
\begin{equation}\label{eq:tr}
    \text{Tr} \left[ J_s\right] =  -\frac{\tilde{c} v_s}{2 \xi_s }+ (1-v_s^2) \left[ \frac{k'(v_s)}{\xi_s}-\left(\frac{4}{n}-m\right)\right],
\end{equation}
and
\begin{equation}\label{eq:det}
    \text{Det}\left[J_s\right]= \frac{\left(1-v_s^2\right) \left[4 \tilde{c}v_s (n-2)  (4-mn)-k'(v_s) \left(v_s^2 (4-mn)-2 (n-2)\right)^2\right]}{2 \tilde{c} n^2 v_s}.
\end{equation}
respectively, where in the second equation we eliminated $\xi_s$ for convenience (through the vanishing of the RHS in \eqref{eq:sys}).

Let begin by discussing the case $ mn < 4$. Since $\mathcal{F}(0) >0, \mathcal{F}(\sqrt{2}/2) <0$ and $\mathcal{F}'(v) <0 $ for $0 \leq v^2 \leq 1/2$, there always exists exactly one solution in the correct range.\footnote{We are also assuming $n>2$.} If $k'(v_s) <0$, as is true for any fixed point with $mn <4$, \eqref{eq:tr} implies $ \text{Tr} \left[ J_s\right] <0$, and from \eqref{eq:det} $ \text{Det}\left[J_s\right]>0$. Therefore, this fixed point is always stable whenever it exists, matching with the expectations for the constant tension case, $m=0$. 

For $mn >4$, however, there are always two solutions to $\mathcal{F}(v)=0$ , as $\mathcal{F}(\sqrt{2}/2),\mathcal{F}(1) >0$, and $\mathcal{F}(v)$ has a single minimum in the interval $1/2 < v^2 <1$. The one closer to $v^2=1/2$ has $k'(v) <0$, while the one closer to $v^2=1$ has $k'(v) >0$.  If $k'(v_s) >0$,  $ \text{Det}\left[J_s\right] <0$, and the latter fixed point can never be stable. For the remaining one to be stable, two conditions need to be satisfied:  $ \text{Det}\left[J_s\right] > 0$, and $ \text{Tr} \left[ J_s\right] <0$. Using the form of $\mathcal{F}(v)$ given by \eqref{eq:f(v)}, these translate to\footnote{Assuming $k(v_s) <0$.}
\begin{equation}
    \frac{k'(v_s)}{k^2(v_s)} < -\frac{4(n-2)}{mn-4}\frac{1}{\tilde{c}v_s^3}
\end{equation}
and
\begin{equation}
    2(1-v_s^2)k(v_s) > v_s \left[ 2(1-v_s^2)k'(v_s)-\tilde{c}v_s\right].
\end{equation}
respectively. It can be verified numerically, that for reasonable values of $\tilde{c}$ and for $ m n >4$, the above conditions are always satisfied with the form of $k(v)$ given in \eqref{eq:k(v)}. Therefore, the scaling fixed point with the lowest value of $v_s^2$ is always stable.

\subsection*{Ultra-relativistic fixed points}

If $\tilde{c}$ asymptotes to a constant value $\tilde{c}_{\infty}$ for $v \rightarrow 1$, there is a trivial fixed point given by
\begin{equation}
    v_r= 1 \quad \quad \quad \xi_r = \frac{\tilde{c}_{\infty}n}{2(n-2)n+mn-4}.
\end{equation}
Of course, $v_r=1$ exactly is inconsistent for a massive string; such a fixed point will never be reached in a finite amount of time, and it is only an asymptotic limit. Moreover, it is possible that additional, higher-order relativistic corrections will  push to $v_r$ to some finite value very close to (but lower) than $v_r=1$. Its stability depends on the eigenvalues of the Jacobian
\begin{equation}\label{eq:jcrel2}
J_r=
    \begin{pmatrix}
        \frac{4}{n}-\frac{m}{2}-1 & \xi_r \left( \frac{4}{n}-m\right)+ \frac{\tilde{c}_{\infty}+\tilde{c}'(1)}{2} \\
        0 & 2 \left( \frac{4}{n}-m\right) \\
    \end{pmatrix},
\end{equation}
where the prime denotes a derivative with respect to $v$. Precisely for $m n >4$ the last eigenvalue changes sign, and the fixed point is stable if (and only if)
\begin{equation}\label{eq:stb}
    mn > {\rm Max} \left\{ 4, 4+n \left(\frac{m}{2}-1 \right)\right\}.
\end{equation}
Notice that this last conclusion doesn't depend on the specific form taken by $\tilde{c}(v)$, as the eigenvalues only depend on the diagonal entries.
Since $\xi_r$ is constant, this is a proper scaling solution, where the correlation length grows linearly with time and $\rho/H^2 \sim \mu (t)$. However, it is instructive to see how the attractor is approached if $\tilde{c}_{\infty} \ll 1$. In the regime where $v_r \simeq 1$, and $\tilde{c}(v) \simeq \tilde{c}_{\infty}$, the evolution equation for the correlation parameter $\xi$ can be solved exactly as
\begin{equation}\label{eq:xiex}
   \xi = \frac{\tilde{c}_{\infty}n}{2(n-2n+mn-4}+  \left[ \xi_0 -\frac{\tilde{c}_{\infty}n}{2(n-2)+mn-4} \right]  \left(\frac{t}{t_0}\right)^{-\frac{1}{2n}\left[ 2(n-2)+ mn -4 \right]},
\end{equation}
where we have reinstated the dependence in terms of $t$ (rather than $z$), and $\xi_0 = \xi(t_0)$ is simply fixed in terms of initial conditions. If $\tilde{c}_{\infty} \ll \xi_0 \sim \mathcal{O}(1)$, there is a long interval on the approach to tracker, lasting until a time $t_r$ satisfying
\begin{equation}
    \frac{t_r}{t_0} \lesssim \left( \frac{\xi_0}{\tilde{c}_{\infty}} \frac{2(n-2)+mn-4}{n} \right)^{\frac{2n}{2(n-2)+mn-4}} ,
\end{equation}
where the constant term in \eqref{eq:xiex} can be neglected, and the energy density in long strings redshifts as radiation:
\begin{equation}
    \rho= \frac{\mu(t)}{L(t)^2} = \frac{\mu(t_0)}{\xi_0^2} \left(\frac{t_0}{t}\right)^{8/n}\sim \frac{1}{a(t)^4}.
\end{equation}
Therefore, even if the actual asymptotic solution is a genuine scaling regime, the approach to the attractor can be characterised by a long phase where the network behaves like radiation. This matches with intuitive expectations for a gas of highly relativistic strings, and can also be seen directly from \eqref{eq:rho1} if $\tilde{c} \ll 1$. A similar behaviour also arises for a cosmic string network in a contracting universe \cite{Avelino:2002hx,Avelino:2002xy}, with the difference that the radiation (as opposed to scaling) regime is the actual endpoint of the evolution in that case, rather than a transient phase.

What happens if $\tilde{c}(v)$ vanishes in the limit $v\rightarrow 1$? Naively, one would like to repeat the previous reasoning, and identify a fixed point with
\begin{equation}
    v_r= 1 \quad \quad \quad \xi_r = 0.
\end{equation}
However, $\xi = 0$ is actually a singular point of the system \eqref{eq:sys}, and some more care should be taken. It will be convenient to reformulate the system close to the line $v=1$, and expand all coefficients to the first non-trivial order in $w \equiv 1-v$. We further assume the chopping efficiency to be given by an appropriate power of the Lorentz contraction factor
$\tilde{c}(v) = \tilde{c}(1-v^2)^{\alpha}$, with $0 <\alpha < 1$. The value $\alpha = \frac{1}{2}$ corresponds to ansatz for $\tilde{c}(v)$ motivated in \cite{Avelino:2002hx,Avelino:2002xy}. Then, \eqref{eq:sys} in the ultra-relativistic regime  
(with $w \ll 1$) can be recast as
\begin{equation}\label{eq:sys1}
    \begin{cases}
        \frac{{\rm d} \xi}{{\rm d}z} = - \frac{2(n-2)+mn-4}{2n} \xi+ \frac{\tilde{c}}{2}(2w)^{\alpha}\\
         \frac{{\rm d} w}{{\rm d}z} = -2\frac{mn-4}{n}w^2 +k'(1) \frac{w^2}{\xi},\\
    \end{cases}
\end{equation}
where $k'(1)\equiv k'(v)\lvert_{v=1}$. With the substitution
\begin{equation}
    \rho= \frac{w^{2-\alpha}}{\xi} \quad \quad \sigma = w^{1-\alpha},
\end{equation}
Eq. \eqref{eq:sys1} can be further simplified to
\begin{equation}\label{eq:sys2}
    \begin{cases}
        \frac{{\rm d} \sigma}{{\rm d}z} = - 2(1-\alpha)\frac{mn-4}{n} \sigma+ k'(1) \rho (1-\alpha)\\
         \frac{{\rm d} \rho}{{\rm d}z} =  \left(\frac{2(n-2)+mn-4}{2n}-2 (2-\alpha) \frac{mn-4}{n}\right)\rho+ k'(1)(2-\alpha)\rho^2 \sigma- \frac{\tilde{c}}{2^{1-\alpha}} \rho^2.\\
    \end{cases}
\end{equation}
This has the advantage of having a completely regular fixed point, $\rho = \sigma =0$, which corresponds to the limit $v \rightarrow 1$ for $ \alpha < 1$. The Jacobian evaluated at that point is
\begin{equation}\label{eq:jcrel}
J_r=
    \begin{pmatrix}
      - 2(1-\alpha)\frac{mn-4}{n}  &  k'(1)(1-\alpha) \\
        0 & \left(\frac{2(n-2)+mn-4}{2n}-2 (2-\alpha) \frac{mn-4}{n}\right) \\
    \end{pmatrix}
\end{equation}
and stability is achieved for $mn > 4$ and 
\begin{equation}
    \alpha < \frac{7(mn-4)-2(n-2)}{4(mn-4)},
\end{equation}
which for $\alpha = \frac{1}{2}$ becomes $m > \frac{16+2n}{5n}$. Incidentally, this is very narrowly satisfied for the volume kination values of $m=1,n=6$. For $v$ close to $1$, the correlation parameter $\xi(t) $ will vanish asymptotically as
\begin{equation}
     \xi(t) = \xi_0 \left(\frac{t}{t_0}\right)^{-\frac{1}{2n}\left[ 2(n-2)+ mn -4 \right]},
\end{equation}
so that the total energy density will red-shift as 
\begin{equation}
    \rho= \frac{\mu(t)}{L(t)^2} \sim \frac{1}{t^{8/n}} \sim \frac{1}{a(t)^4}.
\end{equation}
Thus, at the fixed point will genuinely mimic a radiation fluid. However, as in the case where $\tilde{c}_{\infty} >0$, even when the fixed point above is a saddle there can be long transients characterised by $v \simeq 1$, in which the network will behave as radiation. Since $z = \log t$, these can extend across many orders of magnitude in physical time $t$, and thus behave like a true attractor for many practical purposes.

\section{General momentum parameter}\label{app:k(v)}

In the main text (and the previous appendix), we relied heavily on the explicit for of $k(v)$, which depends on a combination of analytical arguments and numerical interpolations.\footnote{In particular, its functional form has been numerically verified only in the regime $v^2 \leq \frac{1}{2}$.} To make our results more robust, we can also derive bounds which hold for more arbitrary functions $k(v)$, modulo mild assumptions. Assuming a constant $\tilde{c}$, one can rewrite \eqref{eq:L} as
\begin{equation}\label{eq:b1}
    \frac{{\rm d} L}{{\rm d}t}=\frac{2L}{nt} \left[1-v^2\left( \frac{mn}{4}-1\right) \right]+ \frac{\tilde{c} v}{2}.
\end{equation}
If $ m n >4 $ and $k(v) > 0$ for $v^2< \frac{1}{2}$, $v^2 > \frac{1}{2}$ asymptotically. The, \eqref{eq:b1} implies
\begin{equation}\label{eq:quasisc}
   \frac{2 L}{t} \left(\frac{2}{n}- \frac{m}{4} \right)+ \frac{\tilde{c}}{2 \sqrt{2}} < \frac{{\rm d} L}{{\rm d}t} <
   \frac{L}{t} \left(\frac{3}{n}- \frac{m}{4} \right)+ \frac{\tilde{c}}{2},
\end{equation}
which in turn can be integrated to
\begin{equation}
 \frac{\tilde{c}  t}{2\sqrt{2} \left[1-\left(\frac{4}{n}-\frac{m}{2} \right)\right]}+c_1 t^{\frac{4}{n}-\frac{m}{2}} < L(t)-L(0) < \frac{ \tilde{c} n t}{2\left[1-\left(\frac{3}{n}-\frac{m}{4} \right)\right]}+c_2 t^{\frac{3}{n}-\frac{m}{4}}
\end{equation}
where $c_1,c_2$ are arbitrary constants. If the background fluid satisfies $ n>2$ \footnote{Notice how $n=2$ would correspond to a background of non-interacting strings.} (and since $m n >4$), both $\frac{4}{n}-\frac{m}{2} <1$ and $\frac{3}{n}-\frac{m}{4} <1$. Therefore, $L(t)$ is asymptotically bounded between two linear functions, which is a weak form of the scaling regime. 

\bibliography{biblio}

\providecommand{\href}[2]{#2}\begingroup\raggedright\begin{thebibliography}{100}

\bibitem{Kibble:1976sj}
T.~W.~B. Kibble, \emph{{Topology of Cosmic Domains and Strings}},
  \href{https://doi.org/10.1088/0305-4470/9/8/029}{\emph{J. Phys. A} {\bfseries
  9} (1976) 1387}.

\bibitem{Kibble:1984hp}
T.~W.~B. Kibble, \emph{{Evolution of a system of cosmic strings}},
  \href{https://doi.org/10.1016/0550-3213(85)90596-6}{\emph{Nucl. Phys. B}
  {\bfseries 252} (1985) 227}.

\bibitem{Hindmarsh:1994re}
M.~B. Hindmarsh and T.~W.~B. Kibble, \emph{{Cosmic strings}},
  \href{https://doi.org/10.1088/0034-4885/58/5/001}{\emph{Rept. Prog. Phys.}
  {\bfseries 58} (1995) 477}
  [\href{https://arxiv.org/abs/hep-ph/9411342}{{\ttfamily hep-ph/9411342}}].

\bibitem{Vilenkin:2000jqa}
A.~Vilenkin and E.~P.~S. Shellard, \emph{{Cosmic Strings and Other Topological
  Defects}}. Cambridge University Press, 7, 2000.

\bibitem{Copeland:2009ga}
E.~J. Copeland and T.~W.~B. Kibble, \emph{{Cosmic Strings and Superstrings}},
  \href{https://doi.org/10.1098/rspa.2009.0591}{\emph{Proc. Roy. Soc. Lond. A}
  {\bfseries 466} (2010) 623}
  [\href{https://arxiv.org/abs/0911.1345}{{\ttfamily 0911.1345}}].

\bibitem{Charnock:2016nzm}
T.~Charnock, A.~Avgoustidis, E.~J. Copeland and A.~Moss, \emph{{CMB constraints
  on cosmic strings and superstrings}},
  \href{https://doi.org/10.1103/PhysRevD.93.123503}{\emph{Phys. Rev. D}
  {\bfseries 93} (2016) 123503}
  [\href{https://arxiv.org/abs/1603.01275}{{\ttfamily 1603.01275}}].

\bibitem{EPTA:2023xxk}
{\scshape EPTA, InPTA} collaboration, J.~Antoniadis et~al., \emph{{The second
  data release from the European Pulsar Timing Array - IV. Implications for
  massive black holes, dark matter, and the early Universe}},
  \href{https://doi.org/10.1051/0004-6361/202347433}{\emph{Astron. Astrophys.}
  {\bfseries 685} (2024) A94}
  [\href{https://arxiv.org/abs/2306.16227}{{\ttfamily 2306.16227}}].

\bibitem{NANOGrav:2020bcs}
{\scshape NANOGrav} collaboration, Z.~Arzoumanian et~al., \emph{{The NANOGrav
  12.5 yr Data Set: Search for an Isotropic Stochastic Gravitational-wave
  Background}},
  \href{https://doi.org/10.3847/2041-8213/abd401}{\emph{Astrophys. J. Lett.}
  {\bfseries 905} (2020) L34}
  [\href{https://arxiv.org/abs/2009.04496}{{\ttfamily 2009.04496}}].

\bibitem{NANOGrav:2023gor}
{\scshape NANOGrav} collaboration, G.~Agazie et~al., \emph{{The NANOGrav 15 yr
  Data Set: Evidence for a Gravitational-wave Background}},
  \href{https://doi.org/10.3847/2041-8213/acdac6}{\emph{Astrophys. J. Lett.}
  {\bfseries 951} (2023) L8}
  [\href{https://arxiv.org/abs/2306.16213}{{\ttfamily 2306.16213}}].

\bibitem{NANOGrav:2023hvm}
{\scshape NANOGrav} collaboration, A.~Afzal et~al., \emph{{The NANOGrav 15 yr
  Data Set: Search for Signals from New Physics}},
  \href{https://doi.org/10.3847/2041-8213/acdc91}{\emph{Astrophys. J. Lett.}
  {\bfseries 951} (2023) L11}
  [\href{https://arxiv.org/abs/2306.16219}{{\ttfamily 2306.16219}}].

\bibitem{Gorghetto:2018myk}
M.~Gorghetto, E.~Hardy and G.~Villadoro, \emph{{Axions from Strings: the
  Attractive Solution}},
  \href{https://doi.org/10.1007/JHEP07(2018)151}{\emph{JHEP} {\bfseries 07}
  (2018) 151} [\href{https://arxiv.org/abs/1806.04677}{{\ttfamily
  1806.04677}}].

\bibitem{Gorghetto:2020qws}
M.~Gorghetto, E.~Hardy and G.~Villadoro, \emph{{More axions from strings}},
  \href{https://doi.org/10.21468/SciPostPhys.10.2.050}{\emph{SciPost Phys.}
  {\bfseries 10} (2021) 050}
  [\href{https://arxiv.org/abs/2007.04990}{{\ttfamily 2007.04990}}].

\bibitem{Caprini:2018mtu}
C.~Caprini and D.~G. Figueroa, \emph{{Cosmological Backgrounds of Gravitational
  Waves}}, \href{https://doi.org/10.1088/1361-6382/aac608}{\emph{Class. Quant.
  Grav.} {\bfseries 35} (2018) 163001}
  [\href{https://arxiv.org/abs/1801.04268}{{\ttfamily 1801.04268}}].

\bibitem{Auclair:2019wcv}
P.~Auclair et~al., \emph{{Probing the gravitational wave background from cosmic
  strings with LISA}},
  \href{https://doi.org/10.1088/1475-7516/2020/04/034}{\emph{JCAP} {\bfseries
  04} (2020) 034} [\href{https://arxiv.org/abs/1909.00819}{{\ttfamily
  1909.00819}}].

\bibitem{Sousa:2024ytl}
L.~Sousa, \emph{{Cosmic strings and gravitational waves}},
  \href{https://doi.org/10.1007/s10714-024-03293-x}{\emph{Gen. Rel. Grav.}
  {\bfseries 56} (2024) 105}.

\bibitem{Sousa:2013aaa}
L.~Sousa and P.~P. Avelino, \emph{{Stochastic Gravitational Wave Background
  generated by Cosmic String Networks: Velocity-Dependent One-Scale model
  versus Scale-Invariant Evolution}},
  \href{https://doi.org/10.1103/PhysRevD.88.023516}{\emph{Phys. Rev. D}
  {\bfseries 88} (2013) 023516}
  [\href{https://arxiv.org/abs/1304.2445}{{\ttfamily 1304.2445}}].

\bibitem{Gouttenoire:2019kij}
Y.~Gouttenoire, G.~Servant and P.~Simakachorn, \emph{{Beyond the Standard
  Models with Cosmic Strings}},
  \href{https://doi.org/10.1088/1475-7516/2020/07/032}{\emph{JCAP} {\bfseries
  07} (2020) 032} [\href{https://arxiv.org/abs/1912.02569}{{\ttfamily
  1912.02569}}].

\bibitem{Correia:2024cpk}
J.~Correia, M.~Hindmarsh, J.~Lizarraga, A.~Lopez-Eiguren, K.~Rummukainen and
  J.~Urrestilla, \emph{{Scaling density of axion strings in terasite
  simulations}},  \href{https://arxiv.org/abs/2410.18064}{{\ttfamily
  2410.18064}}.

\bibitem{Saikawa:2024bta}
K.~Saikawa, J.~Redondo, A.~Vaquero and M.~Kaltschmidt, \emph{{Spectrum of
  global string networks and the axion dark matter mass}},
  \href{https://doi.org/10.1088/1475-7516/2024/10/043}{\emph{JCAP} {\bfseries
  10} (2024) 043} [\href{https://arxiv.org/abs/2401.17253}{{\ttfamily
  2401.17253}}].

\bibitem{Hindmarsh:2021mnl}
M.~Hindmarsh, J.~Lizarraga, A.~Urio and J.~Urrestilla, \emph{{Loop decay in
  Abelian-Higgs string networks}},
  \href{https://doi.org/10.1103/PhysRevD.104.043519}{\emph{Phys. Rev. D}
  {\bfseries 104} (2021) 043519}
  [\href{https://arxiv.org/abs/2103.16248}{{\ttfamily 2103.16248}}].

\bibitem{Baeza-Ballesteros:2023say}
J.~Baeza-Ballesteros, E.~J. Copeland, D.~G. Figueroa and J.~Lizarraga,
  \emph{{Gravitational wave emission from a cosmic string loop: Global case}},
  \href{https://doi.org/10.1103/PhysRevD.110.043522}{\emph{Phys. Rev. D}
  {\bfseries 110} (2024) 043522}
  [\href{https://arxiv.org/abs/2308.08456}{{\ttfamily 2308.08456}}].

\bibitem{Baeza-Ballesteros:2024otj}
J.~Baeza-Ballesteros, E.~J. Copeland, D.~G. Figueroa and J.~Lizarraga,
  \emph{{Gravitational Wave Emission from Cosmic String Loops, II: Local
  Case}},  \href{https://arxiv.org/abs/2408.02364}{{\ttfamily 2408.02364}}.

\bibitem{Kume:2024adn}
J.~Kume and M.~Hindmarsh, \emph{{Revised bounds on local cosmic strings from
  NANOGrav observations}},  \href{https://arxiv.org/abs/2404.02705}{{\ttfamily
  2404.02705}}.

\bibitem{Hindmarsh:2022awe}
M.~Hindmarsh and J.~Kume, \emph{{Multi-messenger constraints on Abelian-Higgs
  cosmic string networks}},
  \href{https://doi.org/10.1088/1475-7516/2023/04/045}{\emph{JCAP} {\bfseries
  04} (2023) 045} [\href{https://arxiv.org/abs/2210.06178}{{\ttfamily
  2210.06178}}].

\bibitem{Cui:2017ufi}
Y.~Cui, M.~Lewicki, D.~E. Morrissey and J.~D. Wells, \emph{{Cosmic Archaeology
  with Gravitational Waves from Cosmic Strings}},
  \href{https://doi.org/10.1103/PhysRevD.97.123505}{\emph{Phys. Rev. D}
  {\bfseries 97} (2018) 123505}
  [\href{https://arxiv.org/abs/1711.03104}{{\ttfamily 1711.03104}}].

\bibitem{Cui:2018rwi}
Y.~Cui, M.~Lewicki, D.~E. Morrissey and J.~D. Wells, \emph{{Probing the pre-BBN
  universe with gravitational waves from cosmic strings}},
  \href{https://doi.org/10.1007/JHEP01(2019)081}{\emph{JHEP} {\bfseries 01}
  (2019) 081} [\href{https://arxiv.org/abs/1808.08968}{{\ttfamily
  1808.08968}}].

\bibitem{McAllister:2023vgy}
L.~McAllister and F.~Quevedo, \emph{{Moduli Stabilization in String Theory}},
  \href{https://arxiv.org/abs/2310.20559}{{\ttfamily 2310.20559}}.

\bibitem{Cicoli:2023opf}
M.~Cicoli, J.~P. Conlon, A.~Maharana, S.~Parameswaran, F.~Quevedo and
  I.~Zavala, \emph{{String cosmology: From the early universe to today}},
  \href{https://doi.org/10.1016/j.physrep.2024.01.002}{\emph{Phys. Rept.}
  {\bfseries 1059} (2024) 1}
  [\href{https://arxiv.org/abs/2303.04819}{{\ttfamily 2303.04819}}].

\bibitem{Yamaguchi:2005gp}
M.~Yamaguchi, \emph{{Cosmological evolution of cosmic strings with time
  dependent tension}},
  \href{https://doi.org/10.1103/PhysRevD.72.043533}{\emph{Phys. Rev. D}
  {\bfseries 72} (2005) 043533}
  [\href{https://arxiv.org/abs/hep-ph/0503227}{{\ttfamily hep-ph/0503227}}].

\bibitem{Ichikawa:2006rw}
K.~Ichikawa, T.~Takahashi and M.~Yamaguchi, \emph{{Implications of cosmic
  strings with time-varying tension on CMB and large scale structure}},
  \href{https://doi.org/10.1103/PhysRevD.74.063526}{\emph{Phys. Rev. D}
  {\bfseries 74} (2006) 063526}
  [\href{https://arxiv.org/abs/hep-ph/0606287}{{\ttfamily hep-ph/0606287}}].

\bibitem{Cheng:2008ma}
H.-b. Cheng and Y.-q. Liu, \emph{{The Circular loop equation of a cosmic string
  with time-varying tension}},
  \href{https://doi.org/10.1142/S0217732308026340}{\emph{Mod. Phys. Lett. A}
  {\bfseries 23} (2008) 3023}
  [\href{https://arxiv.org/abs/0801.2808}{{\ttfamily 0801.2808}}].

\bibitem{Wang:2012naa}
L.-L. Wang and H.-B. Cheng, \emph{{The evolution of circular loops of a cosmic
  string with periodic tension}},
  \href{https://doi.org/10.1016/j.physletb.2012.05.034}{\emph{Phys. Lett. B}
  {\bfseries 713} (2012) 59} [\href{https://arxiv.org/abs/1206.2095}{{\ttfamily
  1206.2095}}].

\bibitem{Martins:2018dqg}
C.~J. A.~P. Martins, \emph{{Scaling properties of cosmological axion strings}},
  \href{https://doi.org/10.1016/j.physletb.2018.11.031}{\emph{Phys. Lett. B}
  {\bfseries 788} (2019) 147}
  [\href{https://arxiv.org/abs/1811.12678}{{\ttfamily 1811.12678}}].

\bibitem{Emond:2021vts}
W.~T. Emond, S.~Ramazanov and R.~Samanta, \emph{{Gravitational waves from
  melting cosmic strings}},
  \href{https://doi.org/10.1088/1475-7516/2022/01/057}{\emph{JCAP} {\bfseries
  01} (2022) 057} [\href{https://arxiv.org/abs/2108.05377}{{\ttfamily
  2108.05377}}].

\bibitem{Conlon:2024uob}
J.~P. Conlon, E.~J. Copeland, E.~Hardy and N.~S. Gonz\'alez, \emph{{Percolating
  Cosmic String Networks from Kination}},
  \href{https://arxiv.org/abs/2406.12637}{{\ttfamily 2406.12637}}.

\bibitem{Martins:1996jp}
C.~J. A.~P. Martins and E.~P.~S. Shellard, \emph{{Quantitative string
  evolution}}, \href{https://doi.org/10.1103/PhysRevD.54.2535}{\emph{Phys. Rev.
  D} {\bfseries 54} (1996) 2535}
  [\href{https://arxiv.org/abs/hep-ph/9602271}{{\ttfamily hep-ph/9602271}}].

\bibitem{Martins:2000cs}
C.~J. A.~P. Martins and E.~P.~S. Shellard, \emph{{Extending the velocity
  dependent one scale string evolution model}},
  \href{https://doi.org/10.1103/PhysRevD.65.043514}{\emph{Phys. Rev. D}
  {\bfseries 65} (2002) 043514}
  [\href{https://arxiv.org/abs/hep-ph/0003298}{{\ttfamily hep-ph/0003298}}].

\bibitem{us:2024?}
A.~Ghoshal, F.~Revello and G.~Villa, \emph{{Gravitational waves from cosmic
  (super)strings: the case of a time-varying tension}}. In preparation.

\bibitem{Witten:1985fp}
E.~Witten, \emph{{Cosmic Superstrings}},
  \href{https://doi.org/10.1016/0370-2693(85)90540-4}{\emph{Phys. Lett. B}
  {\bfseries 153} (1985) 243}.

\bibitem{Polchinski:1988cn}
J.~Polchinski, \emph{{Collision of Macroscopic Fundamental Strings}},
  \href{https://doi.org/10.1016/0370-2693(88)90942-2}{\emph{Phys. Lett. B}
  {\bfseries 209} (1988) 252}.

\bibitem{Copeland:2003bj}
E.~J. Copeland, R.~C. Myers and J.~Polchinski, \emph{{Cosmic F and D strings}},
  \href{https://doi.org/10.1088/1126-6708/2004/06/013}{\emph{JHEP} {\bfseries
  06} (2004) 013} [\href{https://arxiv.org/abs/hep-th/0312067}{{\ttfamily
  hep-th/0312067}}].

\bibitem{Damour:2004kw}
T.~Damour and A.~Vilenkin, \emph{{Gravitational radiation from cosmic
  (super)strings: Bursts, stochastic background, and observational windows}},
  \href{https://doi.org/10.1103/PhysRevD.71.063510}{\emph{Phys. Rev. D}
  {\bfseries 71} (2005) 063510}
  [\href{https://arxiv.org/abs/hep-th/0410222}{{\ttfamily hep-th/0410222}}].

\bibitem{Jackson:2004zg}
M.~G. Jackson, N.~T. Jones and J.~Polchinski, \emph{{Collisions of cosmic F and
  D-strings}}, \href{https://doi.org/10.1088/1126-6708/2005/10/013}{\emph{JHEP}
  {\bfseries 10} (2005) 013}
  [\href{https://arxiv.org/abs/hep-th/0405229}{{\ttfamily hep-th/0405229}}].

\bibitem{Dvali:1998pa}
G.~R. Dvali and S.~H.~H. Tye, \emph{{Brane inflation}},
  \href{https://doi.org/10.1016/S0370-2693(99)00132-X}{\emph{Phys. Lett. B}
  {\bfseries 450} (1999) 72}
  [\href{https://arxiv.org/abs/hep-ph/9812483}{{\ttfamily hep-ph/9812483}}].

\bibitem{Burgess:2001fx}
C.~P. Burgess, M.~Majumdar, D.~Nolte, F.~Quevedo, G.~Rajesh and R.-J. Zhang,
  \emph{{The Inflationary brane anti-brane universe}},
  \href{https://doi.org/10.1088/1126-6708/2001/07/047}{\emph{JHEP} {\bfseries
  07} (2001) 047} [\href{https://arxiv.org/abs/hep-th/0105204}{{\ttfamily
  hep-th/0105204}}].

\bibitem{Jones:2002cv}
N.~T. Jones, H.~Stoica and S.~H.~H. Tye, \emph{{Brane interaction as the origin
  of inflation}},
  \href{https://doi.org/10.1088/1126-6708/2002/07/051}{\emph{JHEP} {\bfseries
  07} (2002) 051} [\href{https://arxiv.org/abs/hep-th/0203163}{{\ttfamily
  hep-th/0203163}}].

\bibitem{Sarangi:2002yt}
S.~Sarangi and S.~H.~H. Tye, \emph{{Cosmic string production towards the end of
  brane inflation}},
  \href{https://doi.org/10.1016/S0370-2693(02)01824-5}{\emph{Phys. Lett. B}
  {\bfseries 536} (2002) 185}
  [\href{https://arxiv.org/abs/hep-th/0204074}{{\ttfamily hep-th/0204074}}].

\bibitem{Jones:2003da}
N.~T. Jones, H.~Stoica and S.~H.~H. Tye, \emph{{The Production, spectrum and
  evolution of cosmic strings in brane inflation}},
  \href{https://doi.org/10.1016/S0370-2693(03)00592-6}{\emph{Phys. Lett. B}
  {\bfseries 563} (2003) 6}
  [\href{https://arxiv.org/abs/hep-th/0303269}{{\ttfamily hep-th/0303269}}].

\bibitem{Pogosian:2003mz}
L.~Pogosian, S.~H.~H. Tye, I.~Wasserman and M.~Wyman, \emph{{Observational
  constraints on cosmic string production during brane inflation}},
  \href{https://doi.org/10.1103/PhysRevD.68.023506}{\emph{Phys. Rev. D}
  {\bfseries 68} (2003) 023506}
  [\href{https://arxiv.org/abs/hep-th/0304188}{{\ttfamily hep-th/0304188}}].

\bibitem{Kachru:2003sx}
S.~Kachru, R.~Kallosh, A.~D. Linde, J.~M. Maldacena, L.~P. McAllister and S.~P.
  Trivedi, \emph{{Towards inflation in string theory}},
  \href{https://doi.org/10.1088/1475-7516/2003/10/013}{\emph{JCAP} {\bfseries
  10} (2003) 013} [\href{https://arxiv.org/abs/hep-th/0308055}{{\ttfamily
  hep-th/0308055}}].

\bibitem{Burgess:2022nbx}
C.~P. Burgess and F.~Quevedo, \emph{{RG-induced modulus stabilization:
  perturbative de Sitter vacua and improved D3-$ \overline{\mathrm{D}3} $
  inflation}}, \href{https://doi.org/10.1007/JHEP06(2022)167}{\emph{JHEP}
  {\bfseries 06} (2022) 167}
  [\href{https://arxiv.org/abs/2202.05344}{{\ttfamily 2202.05344}}].

\bibitem{Cicoli:2024bwq}
M.~Cicoli, C.~Hughes, A.~R. Kamal, F.~Marino, F.~Quevedo, M.~Ramos-Hamud
  et~al., \emph{{Back to the origins of brane-antibrane inflation}},
  \href{https://arxiv.org/abs/2410.00097}{{\ttfamily 2410.00097}}.

\bibitem{Aharonov:1987ah}
Y.~Aharonov, F.~Englert and J.~Orloff, \emph{{Macroscopic Fundamental Strings
  in Cosmology}},
  \href{https://doi.org/10.1016/0370-2693(87)90935-X}{\emph{Phys. Lett. B}
  {\bfseries 199} (1987) 366}.

\bibitem{Polchinski:2004ia}
J.~Polchinski, \emph{{Introduction to cosmic F- and D-strings}},  in
  \emph{{NATO Advanced Study Institute and EC Summer School on String Theory:
  From Gauge Interactions to Cosmology}}, pp.~229--253, 12, 2004,
  \href{https://arxiv.org/abs/hep-th/0412244}{{\ttfamily hep-th/0412244}}.

\bibitem{Abel:1999rq}
S.~A. Abel, J.~L.~F. Barbon, I.~I. Kogan and E.~Rabinovici, \emph{{String
  thermodynamics in D-brane backgrounds}},
  \href{https://doi.org/10.1088/1126-6708/1999/04/015}{\emph{JHEP} {\bfseries
  04} (1999) 015} [\href{https://arxiv.org/abs/hep-th/9902058}{{\ttfamily
  hep-th/9902058}}].

\bibitem{Frey:2023khe}
A.~R. Frey, R.~Mahanta, A.~Maharana, F.~Muia, F.~Quevedo and G.~Villa,
  \emph{{String thermodynamics in and out of equilibrium: Boltzmann equations
  and random walks}},
  \href{https://doi.org/10.1007/JHEP03(2024)112}{\emph{JHEP} {\bfseries 03}
  (2024) 112} [\href{https://arxiv.org/abs/2310.11494}{{\ttfamily
  2310.11494}}].

\bibitem{Blanco-Pillado:2021ygr}
J.~J. Blanco-Pillado, K.~D. Olum and J.~M. Wachter, \emph{{Comparison of cosmic
  string and superstring models to NANOGrav 12.5-year results}},
  \href{https://doi.org/10.1103/PhysRevD.103.103512}{\emph{Phys. Rev. D}
  {\bfseries 103} (2021) 103512}
  [\href{https://arxiv.org/abs/2102.08194}{{\ttfamily 2102.08194}}].

\bibitem{Ellis:2023tsl}
J.~Ellis, M.~Lewicki, C.~Lin and V.~Vaskonen, \emph{{Cosmic superstrings
  revisited in light of NANOGrav 15-year data}},
  \href{https://doi.org/10.1103/PhysRevD.108.103511}{\emph{Phys. Rev. D}
  {\bfseries 108} (2023) 103511}
  [\href{https://arxiv.org/abs/2306.17147}{{\ttfamily 2306.17147}}].

\bibitem{Arkani-Hamed:1998jmv}
N.~Arkani-Hamed, S.~Dimopoulos and G.~R. Dvali, \emph{{The Hierarchy problem
  and new dimensions at a millimeter}},
  \href{https://doi.org/10.1016/S0370-2693(98)00466-3}{\emph{Phys. Lett. B}
  {\bfseries 429} (1998) 263}
  [\href{https://arxiv.org/abs/hep-ph/9803315}{{\ttfamily hep-ph/9803315}}].

\bibitem{Balasubramanian:2005zx}
V.~Balasubramanian, P.~Berglund, J.~P. Conlon and F.~Quevedo,
  \emph{{Systematics of moduli stabilisation in Calabi-Yau flux
  compactifications}},
  \href{https://doi.org/10.1088/1126-6708/2005/03/007}{\emph{JHEP} {\bfseries
  03} (2005) 007} [\href{https://arxiv.org/abs/hep-th/0502058}{{\ttfamily
  hep-th/0502058}}].

\bibitem{Conlon:2005ki}
J.~P. Conlon, F.~Quevedo and K.~Suruliz, \emph{{Large-volume flux
  compactifications: Moduli spectrum and D3/D7 soft supersymmetry breaking}},
  \href{https://doi.org/10.1088/1126-6708/2005/08/007}{\emph{JHEP} {\bfseries
  08} (2005) 007} [\href{https://arxiv.org/abs/hep-th/0505076}{{\ttfamily
  hep-th/0505076}}].

\bibitem{Conlon:2022pnx}
J.~P. Conlon and F.~Revello, \emph{{Catch-me-if-you-can: the overshoot problem
  and the weak/inflation hierarchy}},
  \href{https://doi.org/10.1007/JHEP11(2022)155}{\emph{JHEP} {\bfseries 11}
  (2022) 155} [\href{https://arxiv.org/abs/2207.00567}{{\ttfamily
  2207.00567}}].

\bibitem{Giddings:2001yu}
S.~B. Giddings, S.~Kachru and J.~Polchinski, \emph{{Hierarchies from fluxes in
  string compactifications}},
  \href{https://doi.org/10.1103/PhysRevD.66.106006}{\emph{Phys. Rev. D}
  {\bfseries 66} (2002) 106006}
  [\href{https://arxiv.org/abs/hep-th/0105097}{{\ttfamily hep-th/0105097}}].

\bibitem{Apers:2022cyl}
F.~Apers, J.~P. Conlon, M.~Mosny and F.~Revello, \emph{{Kination, meet Kasner:
  on the asymptotic cosmology of string compactifications}},
  \href{https://doi.org/10.1007/JHEP08(2023)156}{\emph{JHEP} {\bfseries 08}
  (2023) 156} [\href{https://arxiv.org/abs/2212.10293}{{\ttfamily
  2212.10293}}].

\bibitem{Revello:2023hro}
F.~Revello, \emph{{Attractive (s)axions: cosmological trackers at the boundary
  of moduli space}},  \href{https://arxiv.org/abs/2311.12429}{{\ttfamily
  2311.12429}}.

\bibitem{Apers:2024ffe}
F.~Apers, J.~P. Conlon, E.~J. Copeland, M.~Mosny and F.~Revello, \emph{{String
  Theory and the First Half of the Universe}},
  \href{https://arxiv.org/abs/2401.04064}{{\ttfamily 2401.04064}}.

\bibitem{Conlon:2024hgw}
J.~P. Conlon, \emph{{Out of the dark: WISPs in String Theory and the Early
  Universe}}, \href{https://doi.org/10.22323/1.454.0001}{\emph{PoS} {\bfseries
  COSMICWISPers} (2024) 001}
  [\href{https://arxiv.org/abs/2402.04725}{{\ttfamily 2402.04725}}].

\bibitem{Conlon:2024ene}
J.~P. Conlon, \emph{{String Theory and the Early Universe: Constraints and
  Opportunities}},  in \emph{{58th Rencontres de Moriond on Cosmology}}, 5,
  2024, \href{https://arxiv.org/abs/2405.19118}{{\ttfamily 2405.19118}}.

\bibitem{Apers:2024dtn}
F.~Apers, J.~P. Conlon and M.~Mosny, \emph{{A Note on 4d Kination and
  Higher-Dimensional Uplifts}},
  \href{https://arxiv.org/abs/2409.08049}{{\ttfamily 2409.08049}}.

\bibitem{Conlon:2008cj}
J.~P. Conlon, R.~Kallosh, A.~D. Linde and F.~Quevedo, \emph{{Volume Modulus
  Inflation and the Gravitino Mass Problem}},
  \href{https://doi.org/10.1088/1475-7516/2008/09/011}{\emph{JCAP} {\bfseries
  09} (2008) 011} [\href{https://arxiv.org/abs/0806.0809}{{\ttfamily
  0806.0809}}].

\bibitem{Yamada:2022aax}
M.~Yamada and K.~Yonekura, \emph{{Cosmic F- and D-strings from pure
  Yang\textendash{}Mills theory}},
  \href{https://doi.org/10.1016/j.physletb.2023.137724}{\emph{Phys. Lett. B}
  {\bfseries 838} (2023) 137724}
  [\href{https://arxiv.org/abs/2204.13125}{{\ttfamily 2204.13125}}].

\bibitem{Yamada:2022imq}
M.~Yamada and K.~Yonekura, \emph{{Cosmic strings from pure
  Yang\textendash{}Mills theory}},
  \href{https://doi.org/10.1103/PhysRevD.106.123515}{\emph{Phys. Rev. D}
  {\bfseries 106} (2022) 123515}
  [\href{https://arxiv.org/abs/2204.13123}{{\ttfamily 2204.13123}}].

\bibitem{Dasgupta:1999ss}
K.~Dasgupta, G.~Rajesh and S.~Sethi, \emph{{M theory, orientifolds and G -
  flux}}, \href{https://doi.org/10.1088/1126-6708/1999/08/023}{\emph{JHEP}
  {\bfseries 08} (1999) 023}
  [\href{https://arxiv.org/abs/hep-th/9908088}{{\ttfamily hep-th/9908088}}].

\bibitem{Baumann:2014nda}
D.~Baumann and L.~McAllister, \emph{{Inflation and String Theory}}, Cambridge
  Monographs on Mathematical Physics. Cambridge University Press, 5, 2015,
  \href{https://doi.org/10.1017/CBO9781316105733}{10.1017/CBO9781316105733},
  [\href{https://arxiv.org/abs/1404.2601}{{\ttfamily 1404.2601}}].

\bibitem{Kachru:2003aw}
S.~Kachru, R.~Kallosh, A.~D. Linde and S.~P. Trivedi, \emph{{De Sitter vacua in
  string theory}},
  \href{https://doi.org/10.1103/PhysRevD.68.046005}{\emph{Phys. Rev. D}
  {\bfseries 68} (2003) 046005}
  [\href{https://arxiv.org/abs/hep-th/0301240}{{\ttfamily hep-th/0301240}}].

\bibitem{Frey:2005jk}
A.~R. Frey, A.~Mazumdar and R.~C. Myers, \emph{{Stringy effects during
  inflation and reheating}},
  \href{https://doi.org/10.1103/PhysRevD.73.026003}{\emph{Phys. Rev. D}
  {\bfseries 73} (2006) 026003}
  [\href{https://arxiv.org/abs/hep-th/0508139}{{\ttfamily hep-th/0508139}}].

\bibitem{Ibanez:2012zz}
L.~E. Ibanez and A.~M. Uranga, \emph{{String theory and particle physics: An
  introduction to string phenomenology}}. Cambridge University Press, 2, 2012.

\bibitem{Marchesano:2022qbx}
F.~Marchesano, B.~Schellekens and T.~Weigand, \emph{{D-brane and F-theory Model
  Building}},  12, 2022, \href{https://arxiv.org/abs/2212.07443}{{\ttfamily
  2212.07443}}.

\bibitem{Marchesano:2024gul}
F.~Marchesano, G.~Shiu and T.~Weigand, \emph{{The Standard Model from String
  Theory: What Have We Learned?}},
  \href{https://arxiv.org/abs/2401.01939}{{\ttfamily 2401.01939}}.

\bibitem{Lanza:2020qmt}
S.~Lanza, F.~Marchesano, L.~Martucci and I.~Valenzuela, \emph{{Swampland
  Conjectures for Strings and Membranes}},
  \href{https://doi.org/10.1007/JHEP02(2021)006}{\emph{JHEP} {\bfseries 02}
  (2021) 006} [\href{https://arxiv.org/abs/2006.15154}{{\ttfamily
  2006.15154}}].

\bibitem{Lanza:2021udy}
S.~Lanza, F.~Marchesano, L.~Martucci and I.~Valenzuela, \emph{{The EFT stringy
  viewpoint on large distances}},
  \href{https://doi.org/10.1007/JHEP09(2021)197}{\emph{JHEP} {\bfseries 09}
  (2021) 197} [\href{https://arxiv.org/abs/2104.05726}{{\ttfamily
  2104.05726}}].

\bibitem{Witten:1984dg}
E.~Witten, \emph{{Some Properties of O(32) Superstrings}},
  \href{https://doi.org/10.1016/0370-2693(84)90422-2}{\emph{Phys. Lett. B}
  {\bfseries 149} (1984) 351}.

\bibitem{Svrcek:2006yi}
P.~Svrcek and E.~Witten, \emph{{Axions In String Theory}},
  \href{https://doi.org/10.1088/1126-6708/2006/06/051}{\emph{JHEP} {\bfseries
  06} (2006) 051} [\href{https://arxiv.org/abs/hep-th/0605206}{{\ttfamily
  hep-th/0605206}}].

\bibitem{Conlon:2006tq}
J.~P. Conlon, \emph{{The QCD axion and moduli stabilisation}},
  \href{https://doi.org/10.1088/1126-6708/2006/05/078}{\emph{JHEP} {\bfseries
  05} (2006) 078} [\href{https://arxiv.org/abs/hep-th/0602233}{{\ttfamily
  hep-th/0602233}}].

\bibitem{Cicoli:2012sz}
M.~Cicoli, M.~Goodsell and A.~Ringwald, \emph{{The type IIB string axiverse and
  its low-energy phenomenology}},
  \href{https://doi.org/10.1007/JHEP10(2012)146}{\emph{JHEP} {\bfseries 10}
  (2012) 146} [\href{https://arxiv.org/abs/1206.0819}{{\ttfamily 1206.0819}}].

\bibitem{Reece:2024wrn}
M.~Reece, \emph{{Extra-Dimensional Axion Expectations}},
  \href{https://arxiv.org/abs/2406.08543}{{\ttfamily 2406.08543}}.

\bibitem{Benabou:2023npn}
J.~N. Benabou, Q.~Bonnefoy, M.~Buschmann, S.~Kumar and B.~R. Safdi,
  \emph{{Cosmological dynamics of string theory axion strings}},
  \href{https://doi.org/10.1103/PhysRevD.110.035021}{\emph{Phys. Rev. D}
  {\bfseries 110} (2024) 035021}
  [\href{https://arxiv.org/abs/2312.08425}{{\ttfamily 2312.08425}}].

\bibitem{Arkani-Hamed:2006emk}
N.~Arkani-Hamed, L.~Motl, A.~Nicolis and C.~Vafa, \emph{{The String landscape,
  black holes and gravity as the weakest force}},
  \href{https://doi.org/10.1088/1126-6708/2007/06/060}{\emph{JHEP} {\bfseries
  06} (2007) 060} [\href{https://arxiv.org/abs/hep-th/0601001}{{\ttfamily
  hep-th/0601001}}].

\bibitem{Bennett:1985qt}
D.~P. Bennett, \emph{{The evolution of cosmic strings}},
  \href{https://doi.org/10.1103/PhysRevD.33.872}{\emph{Phys. Rev. D} {\bfseries
  33} (1986) 872}.

\bibitem{Austin:1993rg}
D.~Austin, E.~J. Copeland and T.~W.~B. Kibble, \emph{{Evolution of cosmic
  string configurations}},
  \href{https://doi.org/10.1103/PhysRevD.48.5594}{\emph{Phys. Rev. D}
  {\bfseries 48} (1993) 5594}
  [\href{https://arxiv.org/abs/hep-ph/9307325}{{\ttfamily hep-ph/9307325}}].

\bibitem{Sakellariadou:2004wq}
M.~Sakellariadou, \emph{{A Note on the evolution of cosmic string/superstring
  networks}}, \href{https://doi.org/10.1088/1475-7516/2005/04/003}{\emph{JCAP}
  {\bfseries 04} (2005) 003}
  [\href{https://arxiv.org/abs/hep-th/0410234}{{\ttfamily hep-th/0410234}}].

\bibitem{Avgoustidis:2005nv}
A.~Avgoustidis and E.~P.~S. Shellard, \emph{{Effect of reconnection probability
  on cosmic (super)string network density}},
  \href{https://doi.org/10.1103/PhysRevD.73.041301}{\emph{Phys. Rev. D}
  {\bfseries 73} (2006) 041301}
  [\href{https://arxiv.org/abs/astro-ph/0512582}{{\ttfamily
  astro-ph/0512582}}].

\bibitem{Pourtsidou:2010gu}
A.~Pourtsidou, A.~Avgoustidis, E.~J. Copeland, L.~Pogosian and D.~A. Steer,
  \emph{{Scaling configurations of cosmic superstring networks and their
  cosmological implications}},
  \href{https://doi.org/10.1103/PhysRevD.83.063525}{\emph{Phys. Rev. D}
  {\bfseries 83} (2011) 063525}
  [\href{https://arxiv.org/abs/1012.5014}{{\ttfamily 1012.5014}}].

\bibitem{Avelino:2012qy}
P.~P. Avelino and L.~Sousa, \emph{{Scaling laws for weakly interacting cosmic
  (super)string and p-brane networks}},
  \href{https://doi.org/10.1103/PhysRevD.85.083525}{\emph{Phys. Rev. D}
  {\bfseries 85} (2012) 083525}
  [\href{https://arxiv.org/abs/1202.6298}{{\ttfamily 1202.6298}}].

\bibitem{Sousa:2016ggw}
L.~Sousa and P.~P. Avelino, \emph{{Probing Cosmic Superstrings with
  Gravitational Waves}},
  \href{https://doi.org/10.1103/PhysRevD.94.063529}{\emph{Phys. Rev. D}
  {\bfseries 94} (2016) 063529}
  [\href{https://arxiv.org/abs/1606.05585}{{\ttfamily 1606.05585}}].

\bibitem{Blanco-Pillado:2013qja}
J.~J. Blanco-Pillado, K.~D. Olum and B.~Shlaer, \emph{{The number of cosmic
  string loops}}, \href{https://doi.org/10.1103/PhysRevD.89.023512}{\emph{Phys.
  Rev. D} {\bfseries 89} (2014) 023512}
  [\href{https://arxiv.org/abs/1309.6637}{{\ttfamily 1309.6637}}].

\bibitem{Blanco-Pillado:2017oxo}
J.~J. Blanco-Pillado and K.~D. Olum, \emph{{Stochastic gravitational wave
  background from smoothed cosmic string loops}},
  \href{https://doi.org/10.1103/PhysRevD.96.104046}{\emph{Phys. Rev. D}
  {\bfseries 96} (2017) 104046}
  [\href{https://arxiv.org/abs/1709.02693}{{\ttfamily 1709.02693}}].

\bibitem{Avelino:2002hx}
P.~P. Avelino, C.~J. A.~P. Martins, C.~Santos and E.~P.~S. Shellard,
  \emph{{Topological defects: A Problem for cyclic universes?}},
  \href{https://doi.org/10.1103/PhysRevD.68.123502}{\emph{Phys. Rev. D}
  {\bfseries 68} (2003) 123502}
  [\href{https://arxiv.org/abs/astro-ph/0206287}{{\ttfamily
  astro-ph/0206287}}].

\bibitem{Avelino:2002xy}
P.~P. Avelino, C.~J. A.~P. Martins, C.~Santos and E.~P.~S. Shellard,
  \emph{{Topological defects in contracting universes}},
  \href{https://doi.org/10.1103/PhysRevLett.89.271301}{\emph{Phys. Rev. Lett.}
  {\bfseries 89} (2002) 271301}
  [\href{https://arxiv.org/abs/astro-ph/0211066}{{\ttfamily
  astro-ph/0211066}}].

\bibitem{Copeland:1998na}
E.~J. Copeland, T.~W.~B. Kibble and D.~A. Steer, \emph{{The Evolution of a
  network of cosmic string loops}},
  \href{https://doi.org/10.1103/PhysRevD.58.043508}{\emph{Phys. Rev. D}
  {\bfseries 58} (1998) 043508}
  [\href{https://arxiv.org/abs/hep-ph/9803414}{{\ttfamily hep-ph/9803414}}].

\bibitem{Allahverdi:2020bys}
R.~Allahverdi et~al., \emph{{The First Three Seconds: a Review of Possible
  Expansion Histories of the Early Universe}},
  \href{https://arxiv.org/abs/2006.16182}{{\ttfamily 2006.16182}}.

\bibitem{Moore:2001px}
J.~N. Moore, E.~P.~S. Shellard and C.~J. A.~P. Martins, \emph{{On the evolution
  of Abelian-Higgs string networks}},
  \href{https://doi.org/10.1103/PhysRevD.65.023503}{\emph{Phys. Rev. D}
  {\bfseries 65} (2002) 023503}
  [\href{https://arxiv.org/abs/hep-ph/0107171}{{\ttfamily hep-ph/0107171}}].

\bibitem{Hindmarsh:2017qff}
M.~Hindmarsh, J.~Lizarraga, J.~Urrestilla, D.~Daverio and M.~Kunz,
  \emph{{Scaling from gauge and scalar radiation in Abelian Higgs string
  networks}}, \href{https://doi.org/10.1103/PhysRevD.96.023525}{\emph{Phys.
  Rev. D} {\bfseries 96} (2017) 023525}
  [\href{https://arxiv.org/abs/1703.06696}{{\ttfamily 1703.06696}}].

\bibitem{Correia:2019bdl}
J.~R. C. C.~C. Correia and C.~J. A.~P. Martins, \emph{{Extending and
  Calibrating the Velocity dependent One-Scale model for Cosmic Strings with
  One Thousand Field Theory Simulations}},
  \href{https://doi.org/10.1103/PhysRevD.100.103517}{\emph{Phys. Rev. D}
  {\bfseries 100} (2019) 103517}
  [\href{https://arxiv.org/abs/1911.03163}{{\ttfamily 1911.03163}}].

\bibitem{Martins:2020jbq}
C.~J. A.~P. Martins, P.~Peter, I.~Y. Rybak and E.~P.~S. Shellard,
  \emph{{Generalized velocity-dependent one-scale model for current-carrying
  strings}}, \href{https://doi.org/10.1103/PhysRevD.103.043538}{\emph{Phys.
  Rev. D} {\bfseries 103} (2021) 043538}
  [\href{https://arxiv.org/abs/2011.09700}{{\ttfamily 2011.09700}}].

\bibitem{Auclair:2022ylu}
P.~Auclair, S.~Blasi, V.~Brdar and K.~Schmitz, \emph{{Gravitational waves from
  current-carrying cosmic strings}},
  \href{https://doi.org/10.1088/1475-7516/2023/04/009}{\emph{JCAP} {\bfseries
  04} (2023) 009} [\href{https://arxiv.org/abs/2207.03510}{{\ttfamily
  2207.03510}}].

\bibitem{Aggarwal:2020olq}
N.~Aggarwal et~al., \emph{{Challenges and opportunities of gravitational-wave
  searches at MHz to GHz frequencies}},
  \href{https://doi.org/10.1007/s41114-021-00032-5}{\emph{Living Rev. Rel.}
  {\bfseries 24} (2021) 4} [\href{https://arxiv.org/abs/2011.12414}{{\ttfamily
  2011.12414}}].

\bibitem{Frey:2024jqy}
A.~R. Frey, R.~Mahanta, A.~Maharana, F.~Quevedo and G.~Villa,
  \emph{{Gravitational Waves from High Temperature Strings}},
  \href{https://arxiv.org/abs/2408.13803}{{\ttfamily 2408.13803}}.

\bibitem{Gouttenoire:2021jhk}
Y.~Gouttenoire, G.~Servant and P.~Simakachorn, \emph{{Kination cosmology from
  scalar fields and gravitational-wave signatures}},
  \href{https://arxiv.org/abs/2111.01150}{{\ttfamily 2111.01150}}.

\bibitem{Servant:2023tua}
G.~Servant and P.~Simakachorn, \emph{{Ultrahigh frequency primordial
  gravitational waves beyond the kHz: The case of cosmic strings}},
  \href{https://doi.org/10.1103/PhysRevD.109.103538}{\emph{Phys. Rev. D}
  {\bfseries 109} (2024) 103538}
  [\href{https://arxiv.org/abs/2312.09281}{{\ttfamily 2312.09281}}].

\bibitem{Blasi:2020wpy}
S.~Blasi, V.~Brdar and K.~Schmitz, \emph{{Fingerprint of low-scale leptogenesis
  in the primordial gravitational-wave spectrum}},
  \href{https://doi.org/10.1103/PhysRevResearch.2.043321}{\emph{Phys. Rev.
  Res.} {\bfseries 2} (2020) 043321}
  [\href{https://arxiv.org/abs/2004.02889}{{\ttfamily 2004.02889}}].

\bibitem{Rattazzi:2003ea}
R.~Rattazzi, \emph{{Cargese lectures on extra-dimensions}},  in \emph{{Cargese
  School of Particle Physics and Cosmology: the Interface}}, pp.~461--517, 8,
  2003, \href{https://arxiv.org/abs/hep-ph/0607055}{{\ttfamily
  hep-ph/0607055}}.

\end{thebibliography}\endgroup
\bibliographystyle{JHEP}

\end{document}